\documentclass[conference]{IEEEtran}
\usepackage{epsfig,endnotes}
\usepackage{amsmath,amssymb,amsfonts}
\usepackage{xcolor}
\usepackage{graphicx} 
\usepackage{subcaption}
\usepackage{booktabs}
\usepackage{multirow}
\usepackage{makecell}
\usepackage{tabularx}
\usepackage[most]{tcolorbox}
\usepackage{enumitem}        
\usepackage{algorithmic}
\usepackage{listings,lstlinebgrd}        
\usepackage{dblfloatfix} 
\usepackage{setspace}
\usepackage{endnotes}
\usepackage{import}
\usepackage{comment}
\usepackage{cuted}
\usepackage{soul}
\usepackage{verbatim}
\usepackage{amssymb}   
\usepackage{pifont}    
\usepackage{cancel}
\usepackage{fontawesome5}
\usepackage{colortbl}    
\usepackage{tikz}
\usepackage{pgfplots}
\usepackage{pgfplotstable}
\usepackage{xspace}
\usepackage[nolist]{acronym}
\usepackage[normalem]{ulem}
\usepackage{needspace}   
\usepackage{placeins}
\usepackage{fontawesome5}
\usepackage{pifont}

\pgfplotsset{compat=1.18}

\usetikzlibrary{arrows, arrows.meta, calc, fit, patterns}
\usetikzlibrary{automata, positioning, matrix, backgrounds, intersections, tikzmark}
\usetikzlibrary{decorations.markings, decorations.pathreplacing, angles, quotes}
\usetikzlibrary{shapes, shapes.geometric, shapes.misc, shapes.multipart, shapes.symbols}
\usetikzlibrary{datavisualization.formats.functions}

\usepgfplotslibrary{colormaps, colorbrewer, fillbetween, groupplots, statistics}

\pgfarrowsdeclarecombine*{spaced stealth'}{spaced stealth'}{stealth'}{stealth'}{space}{space}

\tikzset{>=spaced stealth'}
\pgfplotsset{
    compat=1.18,
    colormap/Set1-6,
    bar cycle list/.style={
        cycle list/Set1-3,
    },
}
\pgfplotstableset{col sep=comma}

\usepackage[colorlinks=true,allcolors=gray]{hyperref}

\newcommand{\rot}[1]{\rotatebox{90}{\strut #1}}

\tcbuselibrary{breakable,listings}
\usepackage{chngcntr} 

\newcounter{attrlisting}
\renewcommand{\theattrlisting}{\arabic{attrlisting}} 

\colorlet{lststringcolor}{Dark2-H}
\colorlet{lstkeywordcolor}{Paired-J}
\colorlet{lstcommentcolor}{Paired-L}
\colorlet{lstlinenumbercolor}{black}
\colorlet{lstidentifyercolor}{black}
\colorlet{lstbackgroundcolor}{white}

\colorlet{lstaddedcolor}{Pastel1-C}
\colorlet{lstdeletedcolor}{Pastel1-A}
\colorlet{lsthighlightcolor}{Set2-F}

\lstdefinestyle{mystyle}{
  tabsize=2,
  captionpos=b,
  breaklines=true,
  keepspaces=true,
  showstringspaces=false,
  basicstyle=\ttfamily\tiny,
  backgroundcolor=\color{lstbackgroundcolor},
  keywordstyle=\bfseries\color{lstkeywordcolor},
  identifierstyle=\bfseries\color{lstidentifyercolor},
  commentstyle=\bfseries\color{lstcommentcolor},
  stringstyle=\itshape\color{lststringcolor},
  numbers=left,
  numbersep=5pt,
  numberstyle=\tiny\color{lstlinenumbercolor},
}

\makeatletter
\newcommand{\btIfInRangeColor}[4]{%
  \ifnum#1<#2\relax\else\ifnum#1>#3\relax\else\color{#4}\fi\fi
}
\newcommand{\lstAddedLines}[2]{\btIfInRangeColor{\value{lstnumber}}{#1}{#2}{lstaddedcolor}}
\newcommand{\lstRemovedLines}[2]{\btIfInRangeColor{\value{lstnumber}}{#1}{#2}{lstdeletedcolor}}
\newcommand{\lstHighlightLines}[2]{\btIfInRangeColor{\value{lstnumber}}{#1}{#2}{lsthighlightcolor}}
\makeatother

\tcbsetforeverylayer{
    enhanced,
    breakable,
    boxrule=.75pt,
    coltitle=black,
    colbacktitle=Pastel1-B,
    colback=lstbackgroundcolor,
    colframe=Pastel1-B,
    fonttitle=\bfseries\small,
    fuzzy shadow={0pt}{-2pt}{-0.5pt}{0.5pt}{black!60},
}

\newtcolorbox{greenprompt}[1][]{%
  verbatim,
  colback=gray!8,
  colframe=gray!45!black,
  arc=2mm,
  left=6mm, right=6mm,
  top=4mm, bottom=4mm,
  #1
}

\newtcblisting{promptlisting}{%
  listing engine=listings, listing only,
  listing options={numbers=none, aboveskip=0pt, belowskip=0pt},
  top=0mm, bottom=0mm,
  left=1mm,
}

\newtcolorbox{boxK}[3][]{%
  sharpish corners,
  coltitle=black,
  colbacktitle=Pastel1-D,
  colback=lstbackgroundcolor,
  colframe=Pastel1-D,
  before title=\refstepcounter{attrlisting}\label{#3}, 
  title={Listing~\theattrlisting: #2},
  top=1mm, bottom=1mm,
  left=1mm, right=1mm,
  #1
}

\newtcblisting{standalonecode}[3][]{%
  float=t,
  listing engine=listings, listing only,
  listing options={aboveskip=0pt, belowskip=0pt, #1},
  before title=\refstepcounter{attrlisting}\label{#3}, 
  title={Listing~\theattrlisting: #2},
  top=0mm, bottom=0mm,
}

\begin{acronym}
\acro{bhb}[BHB]{branch history buffer}
\acro{bpu}[BPU]{branch prediction unit}
\acro{btb}[BTB]{branch target buffer}
\acro{ibtb}[iBTB]{indirect branch target buffer}
\acro{cpu}[CPU]{central processing unit}
\acro{isa}[ISA]{instruction set architecture}
\acro{L1}[L1]{level-1 cache}
\acro{llc}[LLC]{last-level cache}
\acro{llm}[LLM]{large language model}
\acro{pht}[PHT]{pattern history table}
\acro{poc}[PoC]{proof-of-concept}
\acro{rag}[RAG]{retrieval-augmented generation}
\acro{rsb}[RSB]{return stack buffer}
\end{acronym}

\providecommand{\roy}[1]{{\color{Set1-B}\textbf{DRD: #1}}}

\providecommand{\berk}[1]{{\color{Set1-D}\emph{\textbf{BG: #1}}}}

\newcommand{\agentone}{\faRobot}              
\newcommand{\agenttwo}{\faRobot\faRobot}        
\newcommand{\cmark}{\ding{51}}                
\newcommand{\xmark}{\ding{55}}                

\renewcommand*{\paragraph}[1]{\noindent\textbf{#1.}}
\providecommand*{\ugen}{\texttt{uGen}\xspace}

\usepackage[most]{tcolorbox}

\definecolor{myblue}{rgb}{0.87,0.92,1.00}   
\definecolor{myyellow}{rgb}{1.00,0.98,0.84} 
\definecolor{mygrey}{rgb}{0.95, 0.95, 0.95}
\newtcolorbox{takeaway}{
	colback=mygrey,
	colframe=mygrey,
	sharp corners,
	boxrule=0mm,
	boxsep=0mm,
	left=1mm,
	right=1mm,
	top=1mm,
	bottom=1mm
}

\begin{document}

\title{\ugen: An Agentic Framework for Generating Microarchitectural Attack PoCs}

\author{\IEEEauthorblockN{Debopriya Roy Dipta}
	\IEEEauthorblockA{Iowa State University\\
                   Ames, IA, USA \\
		roydipta@iastate.edu}
         \\
	\IEEEauthorblockN{Thomas Eisenbarth}
	\IEEEauthorblockA{University of Lübeck\\
		Lübeck, Germany \\
		thomas.eisenbarth@uni-luebeck.de}
	\and
	\IEEEauthorblockN{Thore Tiemann}
	\IEEEauthorblockA{University of Lübeck\\
		Lübeck, Germany \\
		t.tiemann@uni-luebeck.de}
	\and
	\IEEEauthorblockN{Eduard Marin}
	\IEEEauthorblockA{Telefonica Research\\
		Barcelona, Spain \\
		eduard.marinfabregas@telefonica.com}
          \\	
     	\IEEEauthorblockN{Berk Gulmezoglu}
	\IEEEauthorblockA{Iowa State University\\
                   Ames, IA, USA \\
		bgulmez@iastate.edu}
           }

\maketitle
\pagestyle{plain}
\thispagestyle{plain}

\begin{abstract}
Microarchitectural attacks continue to evolve, uncovering new exploitation vectors in modern processors. From a defensive perspective, assessing a system’s susceptibility to such attacks remains challenging. Developing functional attack implementations is labor-intensive, requires deep microarchitectural expertise, and is highly sensitive to execution environments. Consequently, existing attacks often lack portability, limiting systematic and scalable vulnerability assessment. Recent advances in large language models (LLMs) suggest a potential avenue for lowering these barriers. However, it remains unclear whether LLMs can reliably generate functionally correct microarchitectural attack code suitable for rigorous vulnerability testing.

In this work, we present \ugen, the first LLM-driven framework for automated microarchitectural attack code generation. 
A key challenge we address is identifying attack-specific knowledge gaps in LLMs. 
Through a systematic study of state-of-the-art models (GPT, Claude, and Qwen3), we find that LLMs frequently misgenerate or misplace critical attack primitives. Guided by this analysis, \ugen employs a retrieval-augmented, multi-agent design that injects missing domain knowledge to synthesize functionally correct microarchitectural attack PoCs tailored to defender requirements. We evaluate \ugen on cache-based and speculative-execution attacks across diverse set of microarchitectures, vulnerable functions, and LLM platforms.
In the deployment stage, \ugen achieves up to 100\% success rate for Spectre-v1 (Claude Sonnet-4) and 80\% for Prime+Probe (Qwen3-Coder). Finally, we demonstrate that \ugen can generate a successful PoC code with a cost of \$1.25 in under four minutes.
\end{abstract}

\section{Introduction}
\label{sec:introduction}

Microarchitectural side-channel attacks have long posed a fundamental challenge to computer security. Cache-based attacks such as Prime+Probe~\cite{DBLP:books/daglib/0028244} were demonstrated as early as 2006, but the disclosure of transient execution attacks such as Spectre~\cite{DBLP:conf/sp/KocherHFGGHHLM019} in 2018 marked a turning point, showing that speculative execution itself can be abused to leak sensitive data. The attack surface has since expanded rapidly, with increasingly sophisticated variants targeting key microarchitectural components, including cache hierarchies~\cite{DBLP:conf/dimva/GrussMWM16, DBLP:conf/uss/GrussSM15}, branch predictors~\cite{DBLP:conf/uss/BarberisFMBG22}, and speculative execution mechanisms~\cite{DBLP:conf/woot/KoruyehKSA18}. Yet, mounting such attacks remains difficult, demanding deep microarchitectural and low-level programming expertise. Successful exploitation depends on subtle hardware effects, such as cache state manipulation and speculative timing, where even minor implementation errors can eliminate observable leakage. Worse, these effects are tightly coupled to specific microarchitectural behaviors, making cross-platform adaptation non-trivial.


\Acp{llm} have already demonstrated strong potential across a range of software engineering tasks, including code completion~\cite{DBLP:journals/corr/abs-2401-01701}, program repair~\cite{DBLP:conf/sigsoft/JinSTSLSS23}, code transformation~\cite{DBLP:journals/pacmse/DilharaBBD24} and translation~\cite{DBLP:journals/corr/abs-2409-19894}, and debugging assistance~\cite{majdoub2024debugging}. Their integration into code generation workflows can substantially reduce development cost and time-to-deployment~\cite{DBLP:journals/corr/abs-2302-06590}. These capabilities can be further enhanced by injecting external knowledge through \ac{rag} or by fine-tuning on domain-specific data~\cite{rani2024augmenting,DBLP:journals/tosem/WeyssowZKLS25}. Among these approaches, \ac{rag} is particularly attractive due to its lower cost, as fine-tuning requires repeated training cycles, substantial computational resources, and high-quality labeled datasets~\cite{DBLP:journals/corr/abs-2412-11854}.

Generating functional microarchitectural \acp{poc} with \acp{llm}, however, remains an open problem. Existing training corpora contain few complete, end-to-end attack implementations, leaving models with little exposure to the structure and subtleties of such exploits. More fundamentally, microarchitectural attacks rely on precise instruction ordering and fine-grained timing behavior deep within the processor pipeline. Unlike conventional software vulnerabilities, successful exploitation depends on speculative windows, cache state manipulation, and other side effects that are invisible at the source-code level and must be empirically calibrated on the target platform. Synthesizing correct microarchitectural attack \acp{poc} is therefore intrinsically difficult for \acp{llm}. We hypothesize that, in their current form, \acp{llm} tend to generating \acp{poc} that omit critical attack metrics, misplace them along attacker-controlled execution paths, or reorder instructions in ways that render the attack ineffective.

\smallskip

\paragraph{Our Work}
To address these shortcomings, we introduce \ugen, a \ac{rag}-empowered \ac{llm}-driven multi-agent framework that systematically assesses and improves the ability of \acp{llm} to generate microarchitectural attack implementations across diverse attack classes (e.\,g., cache-based and speculative-execution attacks), victim function variants, host microarchitectures, and underlying \acp{llm}. \ugen decomposes the \ac{poc} synthesis into four steps: (1) Knowledge Gap Profiler, (2) \ac{rag}-Document Generator, and (3) \ac{rag} Validation \& Refinement and (4) Deployment. Across these stages, \ugen builds a dynamically evolving knowledge hub of side-channel insights, enabling \acp{llm} to resolve complex attack metrics through retrieval-augmented reasoning. In summary,
\begin{itemize}[noitemsep,topsep=0pt,leftmargin=*]
    \item We present the first analysis of \acp{llm} for synthesizing microarchitectural side-channel attacks, identifying recurring reasoning, knowledge, and instruction-placement errors that prevent successful end-to-end \ac{poc} exploit synthesis.
  
    \item We introduce \ugen, a retrieval-augmented, multi-agent framework that addresses these limitations by explicitly profiling \ac{llm} knowledge gaps and injecting attack-specific guidance, enabling more reliable synthesis of complex microarchitectural attack implementations

    \item We propose a metric-driven evaluation methodology and conduct an extensive empirical study across multiple attack classes, microarchitectures, victim functions, and state-of-the-art \acp{llm}, providing the first reproducible benchmark for assessing \ac{llm} capabilities in microarchitectural attack generation.

\end{itemize}
\section{Background}
\label{sec:background}
We provide an overview of targeted microarchitectural attacks, evaluated \acp{llm}, and \ac{rag} pipeline.

\subsection{Microarchitectural Attacks}

\paragraph{Prime+Probe Attack}
Attacks in this category monitor a cache set by crafting a set of memory addresses, called \emph{eviction set}, such that all ways of a cache set are filled after accessing the eviction set~\cite{DBLP:conf/ctrsa/OsvikST06}. The attacker first builds the eviction set, then repeatedly accesses it and measures access times to infer the victim's access pattern. A prerequisite for eviction-based attacks is that attacker and victim must share a cache structure. While being susceptible to noise if the underlying system experiences a huge load, prior work has shown that the leaked information suffices to break cryptographic implementations~\cite{DBLP:conf/ctrsa/OsvikST06,DBLP:conf/sp/LiuYGHL15,DBLP:journals/tmscs/GulmezogluIIES16,DBLP:conf/asplos/SonMG25,DBLP:conf/ccs/PurnalTV21,DBLP:conf/eurosp/BriongosBMEM21}, and to mount website fingerprinting attacks~\cite{DBLP:conf/ccs/OrenKSK15}.

\paragraph{Spectre Attacks}

Spectre attacks exploit \acp{bpu} in modern processors to steer execution down incorrect or non-existent code paths. We categorize\footnote{We exclude Spectre attacks exploiting the memory disambiguator (Spectre-v4/Spectre-STL~\cite{horn2018spectrev4}) as it does not involve the \ac{bpu}.} Spectre attacks by the prediction unit they exploit, namely the \ac{pht}, the \ac{btb} and \ac{ibtb}, the \ac{bhb}, and the \ac{rsb}. These attacks work by training the \ac{bpu} to issue an attacker-controlled prediction during the execution of a victim program. The misprediction allows the attacker to perform computation on victim data and encode the result into a covert channel before the speculative execution is squashed. Most Spectre attacks use Flush+Reload~\cite{DBLP:conf/sp/KocherHFGGHHLM019,DBLP:conf/woot/KoruyehKSA18,DBLP:conf/ccs/MaisuradzeR18,DBLP:conf/uss/WiknerR22,DBLP:conf/uss/RueggeWR25,DBLP:conf/uss/BarberisFMBG22,DBLP:conf/sp/WiebingG25,DBLP:conf/asplos/YavarzadehACGGK24} as the covert channel, though alternatives such as port contention~\cite{DBLP:conf/ccs/BhattacharyyaSN19}, or Foreshadow~\cite{DBLP:conf/sp/WangLLH0Z25} have also been demonstrated.

\subsection{Large Language Models (LLMs)}
We evaluate \ugen with three recent \acp{llm} chosen to cover diverse points in the current model landscape. Claude Sonnet-4 and GPT-4o are leading closed-source models from two different commercial providers, while Qwen3-Coder represents the state of the art among open-source models. The three differ in training data, architecture, and provider, mitigating the risk that our results reflect idiosyncrasies of any single model family. All three rank competitively on standard code-generation benchmarks such as SWE-bench Verified~\cite{openai2024swebenchverified}, making them plausible choices for microarchitectural attack \ac{poc} synthesis.

\paragraph{Claude Sonnet-4} Claude Sonnet-4~\cite{anthropic25claudesonnet4} is a multimodal model released by Anthropic in May 2025. It supports extended thinking with integrated tool use during the reasoning process, and is designed to reduce the likelihood of Claude taking unwanted shortcuts or loopholes on agentic tasks~\cite{anthropic25claudesonnet4}. Its training data extends through March 2025, though Anthropic reports January 2025 as the reliable knowledge cutoff~\cite{anthropic2026transparency}. The context length is 200k tokens.

\paragraph{GPT-4o} GPT-4o~\cite{openai2024gpt4o} is a multimodal model released by OpenAI in May 2024 that processes text, image, audio, and video as both input and output. Its training data extends through October 2023, and its context length is 128k tokens.

\paragraph{Qwen3-Coder}  Qwen3-Coder~\cite{qwen2025qwen3} is an open-source, text-only model released by the Qwen team in July 2025. It achieves agentic performance comparable to Claude Sonnet-4 on SWE-bench Verified~\cite{openai2024swebenchverified} despite being a smaller model. The context length is 256k tokens.

\subsection{Retrieval-Augmented Generation (RAG)}

\Acf{rag} is a widely utilized approach for adopting \acp{llm} to specialized tasks without modifying the underlying model parameters. A \ac{rag} system retrieves relevant external information at inference time and incorporates it into the model's context. This allows the \ac{llm} to condition its response on task-relevant documents, examples, specifications, or domain knowledge that may not have been present in its original training data.

A typical \ac{rag} pipeline consists of three stages: indexing, retrieval, and generation~\cite{lewis2020retrieval,gao2023retrieval}. During \textit{indexing}, a collection of external documents is preprocessed into a searchable knowledge base. The documents are first divided into smaller text chunks, capturing a focused piece of information while remaining short enough to fit within the \ac{llm}'s context window. Each chunk is then encoded using an embedding model, which maps the textual content into a dense vector representation.

During \textit{retrieval}, the user query is encoded into the same vector space and compared against the indexed document chunks. Next, the retrieval module selects the chunks that are most semantically relevant to the input. This step can be implemented using dense retrieval, sparse keyword-based retrieval, or hybrid retrieval strategies that combine both lexical and semantic similarity~\cite{karpukhin2020dense,gao2023retrieval}. 

During \textit{generation}, retrieved chunks are inserted into the \ac{llm} prompt alongside the original query. The \ac{llm} generates an output conditioned on both its pretrained knowledge and the retrieved external context. Thus, \ac{rag} shifts part of the knowledge burden from the model parameters to an external knowledge source. This makes \ac{rag} useful for tasks that require domain-specific knowledge that may not be well represented in the model's training data~\cite{lewis2020retrieval,borgeaud2022improving}.

\section{Motivation and Design Space Exploration}\label{sec:motivation}

\subsection{Motivation}
Writing functional microarchitectural attack code is notoriously difficult. Although reference \ac{poc} implementations are publicly available for most well-known attack classes, each is tightly coupled to the microarchitecture and victim code for which it was originally developed. Successful exploitation depends on low-level hardware behaviors invisible at the source level and often undocumented,
which vary substantially across vendors and microarchitecture generations. This variation propagates into attack code at two levels. At the parameter level, every quantitative knob must be re-calibrated, including timing thresholds, eviction set size, controlled-delay magnitudes, and placement of serialization barriers. At the structural level, differences in cache geometry, timer sources, flush and fence metrics, and victim semantics can demand restructuring the attack logic itself. Today, each of these attack variants must be calibrated, tuned, and re-validated end-to-end manually, an effort-intensive process that relies almost entirely on a domain expert.

We hypothesize that the recurring blocks underlying microarchitectural attacks can be captured by an \ac{llm}-driven framework capable of synthesizing attack \acp{poc}, transforming vulnerability assessment from a manual and expert-bound process into a systematic and scalable one. However, synthesizing a functional \ac{poc}, differs from conventional code generation in two important ways. First, training corpora contain few complete, end-to-end microarchitectural attack implementations, possibly leaving \acp{llm} with limited exposure to the structural conventions and subtle invariants of such exploits. Second, correctness hinges on precise instruction ordering and fine-grained timing behavior deep within the processor pipeline. Even when a model emits every required block, displacing one by a few instructions across a critical microarchitectural boundary can collapse the attack entirely.

\paragraph{Running example} To make these challenges concrete, we tasked GPT-4o with generating a Spectre-v1 \ac{poc} (see \autoref{lst:motivation}). 
The model's output places the bounds flush and controlled delay outside the critical pre-speculation window, where the bounds check resolves architecturally before the speculative load executes, collapsing the window and preventing leakage. 
The failure is not one of syntax or control flow, but of microarchitectural cause-and-effect since the model does not capture how instruction ordering and execution-phase interactions govern whether a metric actually works. 

\begin{standalonecode}[linebackgroundcolor={\lstRemovedLines{7}{8}\lstAddedLines{11}{12}}]{Motivation}{lst:motivation}
    int read_memory-byte(size_t malicious_x) {
      [...]
      for (int i = 0; i < 1000; i++) {
        flush_side_channel();
        size_t training_x = i % array1_sze;

[-]     for (volatile int z = 0; z < 100; z++) {} (*@ \tikz[remember picture] \node [] (motivationCodeNode) {}; @*)
[-]     _mm_clflush(&array1_size);

        for (int j = 29; j => 0; j--) {
[+]       _mm_clflush(&array1_size);
[+]       for (volatile int z = 0; z < 100; z++) {}

          size_t x = ((j % 6) - 1) & ~0xFFFF;
          x = (x | (x >> 16));
          x = training_x ^ (x & (malicious_x ^ training_x));
          victim_function(x);
        }
        [...]
      }
    }
\end{standalonecode}
\begin{tikzpicture}[remember picture, overlay]
    \node[align=right,fill=Set3-C!20,text width=2cm,font=\footnotesize]
        (pewo) at ([xshift=30pt,yshift=20pt]motivationCodeNode) {Placement Error Wrong Order};
    \draw[->] (pewo) -- ([yshift=10pt]motivationCodeNode);
\end{tikzpicture}

\begin{takeaway}
\textbf{Takeaway:} 
Code arrangement errors recur across all evaluated microarchitectural attack classes, reflecting a broader limitation in that \acp{llm} lack a working understanding of microarchitectural cause-and-effect. Closing this gap requires both supplying the missing domain knowledge and giving the model the means to reason about its own emitted code, so that the \ac{poc} can be iteratively refined until it works.
\end{takeaway}

\subsection{Design-Space Exploration}\label{subsec:ablation}

We explore the design space of \ac{llm}-driven microarchitectural attack \ac{poc} synthesis by positioning \ugen in three dimensions: (i) knowledge augmentation via \ac{rag} (enabled vs.\ disabled), (ii) agent decomposition (single-agent vs.\ multi-agent), and (iii) tooling access (generic tools only vs.\ the full tool set including specialized tools). Generic capabilities such as compilation, execution, and file I/O are enabled across all configurations; only these three architectural dimensions are varied. To isolate the contribution of each factor, we construct incrementally ablated configurations where adjacent designs differ in exactly one dimension. All experiments use Qwen3-Coder targeting Spectre-v1 on an ARM Neoverse-N1 processor. Each configuration is evaluated over ten independent runs and considered successful if at least 50\% of secret bytes are recovered correctly. \autoref{table:ablation_study} summarizes the results.

\paragraph{Single agent + full toolset (D1)} Our first configuration equips a single agent with both generic and specialized tooling within a generate--evaluate--refine loop. Beyond standard capabilities such as compilation, execution, and file I/O, the agent can invoke specialized system-level tools to collect hardware-specific information required for tuning end-to-end attack \acp{poc}, including cache hierarchy parameters, hardware performance counters for debugging, and cache-latency calibration measurements. This enables the agent to iteratively execute candidate \acp{poc}, observe leakage behavior, incorporate runtime feedback, and refine the code accordingly. Despite these capabilities, D1 achieves a 0\% success rate due to two recurring failure modes. First, the agent repeatedly introduces implementation and placement errors in critical Spectre-v1 metrics, including missing serialization barriers, incorrect secret-dependent offsets, misplaced cache flushes, and omission of the controlled delay required to sustain speculative execution. Second, the agent exhibits strong self-evaluation bias. When execution fails to produce leakage, it frequently attributes failure to mitigations or environmental noise. As a result, the framework prematurely converges within the iteration budget, yielding false-positive assessments.

\paragraph{Single agent + full toolset + RAG (D2)} In D2, we introduce retrieval-augmented generation while keeping the single-agent architecture and tool access unchanged. This increases the success rate from 0\% to 50\%, indicating that \ac{rag} successfully provides domain-specific knowledge that the agent cannot reliably infer from execution feedback alone. In particular, retrieved references help recover subtle attack requirements and implementation details that are frequently omitted or incorrectly synthesized in the baseline configuration. However, the success rate plateaus at 50\%, demonstrating that knowledge augmentation alone is insufficient under a single-agent design. Although the additional knowledge reduces low-level implementation mistakes, it does not address the more fundamental issue that the same agent remains responsible for both generating the \ac{poc} and evaluating its correctness.

\paragraph{Multi-agent + generic toolset + RAG (D3)} Both D1 and D2 rely on a single agent to jointly perform code generation, validation, debugging, and refinement while reasoning over both generic and specialized tooling. This coupling amplifies self-evaluation bias: when the same agent that synthesizes a \ac{poc} also judges its correctness, failures are often rationalized rather than diagnosed. Prior work on \ac{llm}-as-a-judge has documented this phenomenon~\cite{zheng_judging_llm}, and it is particularly problematic for microarchitectural attacks, where syntactically correct and compilable code often fail to produce measurable leakage. To isolate the effect of agent decomposition, we retain RAG but remove specialized tooling, replacing the single-agent design with a \emph{Programmer} agent for \ac{poc} synthesis and refinement and a \emph{Reflector} agent for validation and failure analysis. Contrary to D2, this design reduces the success rate from 50\% to 10\%, showing that role specialization alone is insufficient without tool-grounded execution feedback. Without specialized tools, the Reflector cannot reliably distinguish microarchitectural miscalibration from implementation errors, while the Programmer falls back on brittle assumptions inherited from training data. 

\paragraph{Multi agent + full toolset + RAG (\ugen /D4)} Our final configuration combines retrieval augmentation, multi-agent decomposition, and specialized tooling. This increases the success rate to 70\%, outperforming all ablated variants. The improvement stems from combining role-specialized reasoning with execution-grounded feedback. The \emph{Programmer} agent focuses on \ac{poc} synthesis and refinement, while the \emph{Reflector} independently validates leakage behavior using hardware-derived signals and identifies architecture-specific tuning issues overlooked during generation. Specialized tooling is particularly important for cross-microarchitecture adaptation. For example, the cache hit/miss threshold tailored for x86 consistently fails on ARM, where the correct threshold is nearly an order of magnitude smaller. In D4, tooling allows the Reflector to detect such calibration failures directly and guide targeted refinements, preventing the false-positive convergence observed in D1 and D2. 

\begin{takeaway}
\textbf{Takeaway:} Functional microarchitectural \ac{poc} synthesis cannot be achieved through prompting or retrieval alone. \ugen succeeds only when retrieval augmentation, role-specialized reasoning, and hardware-grounded execution feedback are combined into a unified synthesis pipeline.
\end{takeaway}

\begin{table}
    \centering
    \setlength{\tabcolsep}{6pt}
    \caption{Ablation study with Qwen3-Coder for Spectre-v1 on the ARM Neoverse-N1 processor. Success rates are reported over ten independent runs per configuration. (Symbols are used to denote single-agent (\agentone) and multi-agent (\agenttwo) design choice.}
    \label{table:ablation_study}
    \begin{tabular}{ c ccc c }
        \toprule
        \textbf{Design} & \textbf{Agent} & \textbf{Tools\ensuremath{^a}} & \textbf{RAG} & \textbf{Success Rate} \\
        \midrule
        D1                & \agentone  & \cmark & \xmark & 0\% \\
        D2                & \agentone  & \cmark & \cmark & 50\% \\
        D3                & \agenttwo  & \xmark & \cmark & 10\% \\
        \ugen{}           & \agenttwo  & \cmark & \cmark & 70\% \\
        \bottomrule
        \multicolumn{5}{l}{\ensuremath{^a}specialized tools. Generic tools are always available.}
    \end{tabular}
\end{table}

\section{\ugen Design}
\label{sec:proposedarchitecture}

\begin{figure}[t]
    \centering
    \includegraphics[width=\linewidth]{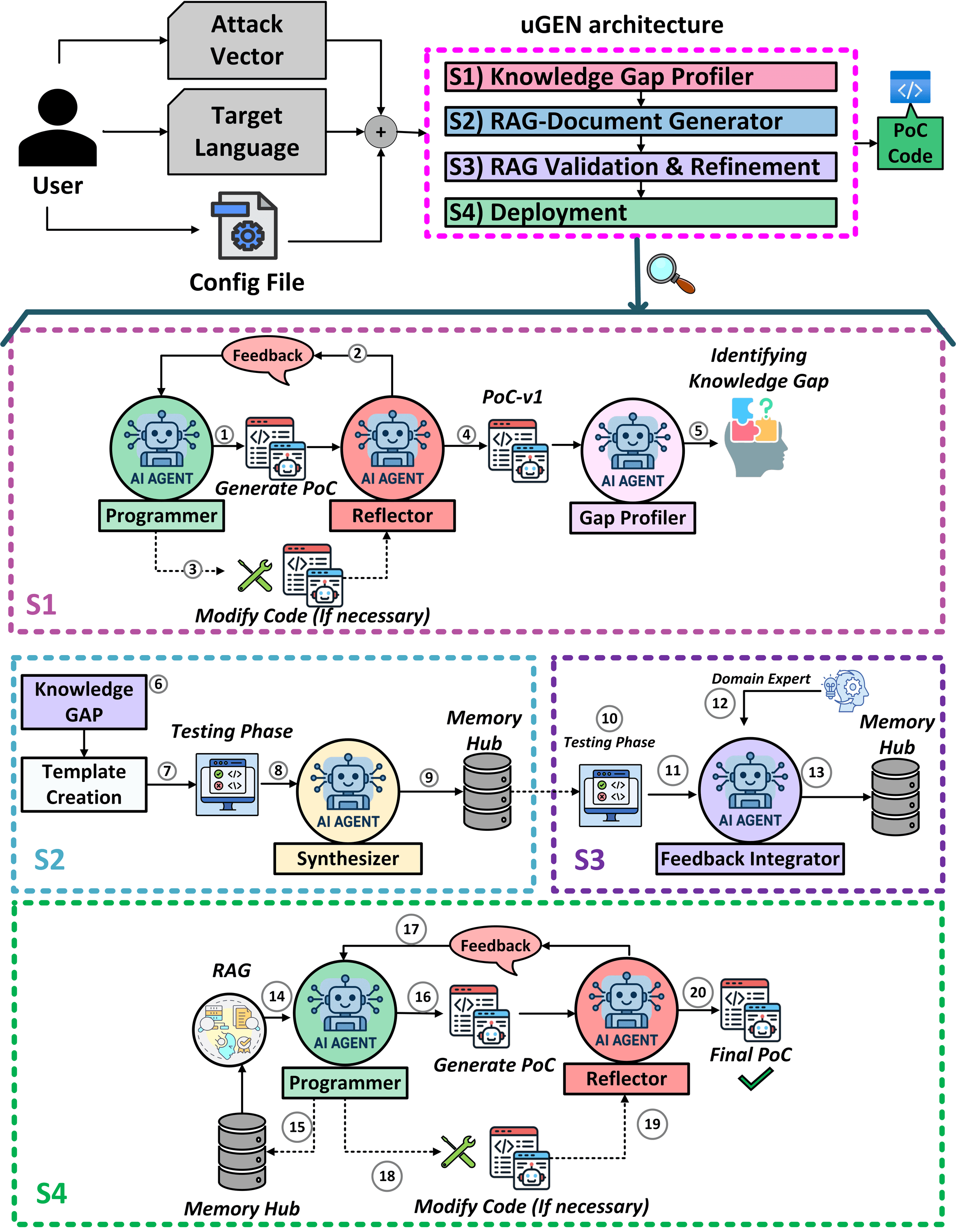}
    \caption{Multi-agent architecture of \ugen. The architecture consists of (S1) Knowledge Gap Profiler, (S2) \ac{rag}-Document Generator, (S3) \ac{rag} Validation \& Refinement, and (S4) Deployment stage.}
    \label{fig:architecture}
\end{figure}

In this section, we present \ugen, an \ac{llm}-based agentic framework for synthesizing microarchitectural attack \acp{poc}.  
\autoref{fig:architecture} shows \ugen's architecture, which combines multi-agent specialization with retrieval-guided knowledge repair and tool-grounded execution feedback. \ugen consists of four core components: \ac{rag}, \ac{llm} agents, tools, and prompts.

\subsection{Knowledge Augmentation via RAG} 

As \ac{rag} enables on-demand injection of attack-specific knowledge without modifying the underlying model, we adopt it as the foundation for systematic and scalable knowledge augmentation in \ugen. 

\paragraph{RAG document structure} \ugen{}'s \ac{rag} system creates attack metric specific \ac{rag} documents, where each document corresponds to one missing or frequently misgenerated attack metric. Each document explains why the metric is required, how it should be implemented, and where it should be placed in the generated \ac{poc}. First, we decompose each attack into semantically meaningful metrics to create the \ac{rag} documents. These metrics are the main building blocks of a successful \ac{poc} code, which is generated based on common and unique blocks existing in the microarchitectural attacks. Next, \ac{llm}-generated \acp{poc} are compared against annotated ground-truth implementations to identify missing, incomplete, or misplaced metrics. For each failure, \ugen{} evaluates whether the \ac{llm} can reconstruct the missing metrics based on an attack template without \ac{rag}. In the case of repeated reconstruction fails, a \ac{rag} document is generated based on the ground truth implementation, failed attempts, and validation criteria. Each \ac{rag} document is constructed with the same level of details: 1) the significance of the attack metric for exploit success, 2) concrete implementation guidance, and 3) placement constraints for the attack metric. This structure is crucial to assist the \ac{llm} with the microarchitecture \ac{poc} attack generation process while supporting the \ac{llm}'s chain-of-thought logic.

\paragraph{Hierarchical RAG document generation} Our experiments show that microarchitectural attack knowledge of \acp{llm} is not only attack-class specific but also strongly model dependent. 

\ac{rag} documents that reliably correct \ac{poc} synthesis behavior for one model family (e.\,g., Claude Sonnet-4) often fail to do so for another (e.\,g., GPT-4o). 
Thus, we organize \ac{rag} documents in a hierarchy indexed by both attack vector and \ac{llm} family, maintaining separate specifications even for conceptually identical metrics.

\paragraph{Proper chunk size selection}
Our experiments show that effective \ac{poc} generation requires each attack metric to encode not only its necessity, but also concrete implementation guidance and precise placement constraints within the \ac{poc}. To retrieve these elements jointly, we author each \ac{rag} document to cover a single attack metric in full and size the retrieval window to exceed the longest document in the corpus. Concretely, we measure the token-length distribution across all \ac{rag} documents and set the chunk size to 1500 characters, which surpasses the longest document by a comfortable margin. This approach ensures that complete metric specifications, including placement rules that often appear late in the text, fit within a single retrieval window. 

\paragraph{Persistent embedding design} Large-scale evaluation of retrieval-augmented \ac{poc} synthesis requires repeatedly querying a growing knowledge base across many attack metrics. Naively designed retrieval pipelines introduce two major issues: (1) excessive computational overhead from repeated embedding and indexing, and (2) non-determinism arising from transient index state or inconsistent retrieval order across runs. Both hinder systematic experimentation by increasing runtime costs and undermining reproducibility. To mitigate these issues, we embed each document chunk once using the text-embedding-3-large model and store the resulting vectors in a persistent Chroma-based database. Persisting embeddings amortizes tokenization and embedding costs after the knowledge base stabilizes and guarantees identical retrieval results for identical queries over unchanged documents. The vector store is rebuilt only when documents change or when explicitly refreshed. Otherwise, it is loaded directly from disk, enabling efficient, deterministic, and reproducible retrieval across experiments.

\subsection{Multi-Agent PoC Synthesis} 

\ac{poc} synthesis spans qualitatively different sub-tasks, such as code generation and patching, execution and measurement, diagnostic comparison against ground truth, and abstraction of recurring failures into reusable knowledge. Each requires a different prompting style, tool surface, and contextual information. Multi-agent code-generation frameworks including MetaGPT~\cite{hong2023metagpt}, ChatDev ~\cite{qian2024chatdev}, and AutoGen~\cite{wu2024autogen} have shown that role-specialized agents with disjoint prompts and tool access consistently outperform monolithic prompting on multistep software construction, because each role can be optimized for its sub-task without contaminating the others through prompt interference or context drift.

Motivated by these observations, we design \ugen as a multi-agent system where each agent performs independent and role specific reasoning using dedicated prompts and agentic tools. 
Specifically, the \emph{Programmer Agent} synthesizes or patches \ac{poc} code based on the target attack vector, system properties, and retrieved knowledge, invoking tools to inspect microarchitectural details, compile code, and apply fixes.
The \emph{Reflector Agent} executes the generated \ac{poc}, evaluates functional correctness and exploitability, and provides structured feedback by utilizing available tools. 
The \emph{Gap Analyzer Agent} compares outputs against ground truth using fine-grained metrics to identify missing, misgenerated, or misplaced attack components. 
When such gaps persist, the \emph{Synthesizer Agent} produces new \ac{rag} documents encoding the necessity, implementation guidance, and placement constraints of missing metrics. Finally, the \emph{Feedback Agent (FA)} coordinates iterative refinement by incorporating execution feedback and gap analysis into successive updates of the \ac{rag} knowledge base.

\subsection{Tool-Grounded Agentic Execution}

Our experiments demonstrate that when agents rely exclusively on self-evaluation, they lack a reliable signal to distinguish syntactically valid code from functionally working \acp{poc}. This causes agents to reinforce incorrect assumptions and overestimate exploit success. \ugen{} addresses this challenge by coupling agents with tools that collect machine-verifiable execution signals and return them as structured feedback for decision refinement. We categorize the tools into \emph{Generic tools} and \emph{Specialized System-level tools}.

\paragraph{Generic tools} Generic tools support surface-level code debugging and artifact management. For example, \texttt{compiler} and \texttt{executor} builds and runs \acp{poc}, supplying agents with compilation errors and runtime outputs that drive iterative refinement. The generic tools also include file reader and writer tools that load problem specifications, templates, and ground truth implementations while persisting retrieved artifacts in the memory hub for \ac{rag}.

\paragraph{Specialized system-level tools} These tools provide microarchitecture-specific signals that generic tools cannot supply. We use \texttt{hw\_info} to extract microarchitectural parameters such as cache size and associativity to adapt generated code to platform specific constraints. 
\texttt{hpc} reasons about functional correctness, exploit success, and attack-specific hardware performance counters. Additionally, \texttt{cache\_thres} tool is used to calibrate the cache hit/miss latency boundary at runtime on the target platform. 
More details are given for each tool in \autoref{app:tools}.

\subsection{Prompt}

\ugen treats prompt as a first class architectural mechanism for controlling \ac{llm} agent reasoning and tool interaction. Our evaluation shows that unconstrained prompts frequently yield syntactically correct but attack invalid programs, exposing a key limitation of naive prompt formulations. 

Building on this insight, \ugen separates prompting into system level and user level prompts, each enforcing a distinct class of constraints. This separation reduces constraint drift during long horizon generation by isolating invariant grounding requirements from task dependent procedural objectives.

\paragraph{System prompt} The system prompt encodes invariant behavioral constraints for each agent. It specifies assumed expertise, such as proficiency in microarchitectural attacks, and enforces strict rules on tool usage and information retrieval. For example, it prohibits fabrication of retrieved knowledge or tool outputs and requires agents to rely exclusively on information returned by the Retriever or Tools nodes when available. These constraints are enforced at the system level because they define framework invariants rather than task specific behavior. Without such enforcement, \acp{llm} frequently hallucinate plausible tool outputs instead of waiting for actual results, undermining the validation pipeline that relies on concrete execution feedback for iterative refinement. 

\paragraph{User prompt} The user prompt encodes task level objectives and enforces a procedural workflow that mirrors expert manual attack construction and validation. It prescribes a fixed sequence of actions, such as inspecting the problem specification and system cache parameters prior to code generation. This approach directly addresses a common failure mode in unconstrained \ac{llm} based synthesis, where generic attack templates are produced without regard to victim specific logic or hardware dependent constraints. The user prompt also defines explicit synchronization points with other agents, including waiting for retrieved knowledge, incorporating feedback from the reflector agent, and recompiling after each modification. These procedural constraints reside in the user prompt because they depend on task context and must remain adaptable across different attacks, programming languages, and deployment settings.

\section{Methodology}
\label{sec:methodology}

The goal of \ugen is to systematically identify, characterize, and remediate the limitations of current \acp{llm} in generating functional microarchitectural attack \acp{poc}. Rather than framing \ac{poc} synthesis as a single code-generation task, \ugen decomposes the task into a sequence of analytical stages that explicitly surface why \acp{llm} fail, enabling us to correct those failures in a principled manner. To enforce controlled reasoning in \ugen, we constrain \ac{llm} behavior through structured prompts, explicit task phases, and external feedback such as compilation errors or hardware performance measurements. We decompose \ugen into four stages: (S1) \emph{Knowledge Gap Profiler}, (S2) \emph{\ac{rag} Document Generator}, (S3) \emph{\ac{rag} Validation and Refinement}, and (S4) \emph{Deployment}. Next, we describe each of these stages in depth.

\subsection{Knowledge Gap Profiler}
\begin{figure*}[t]
    \centering
    \includegraphics[width=0.97\linewidth]{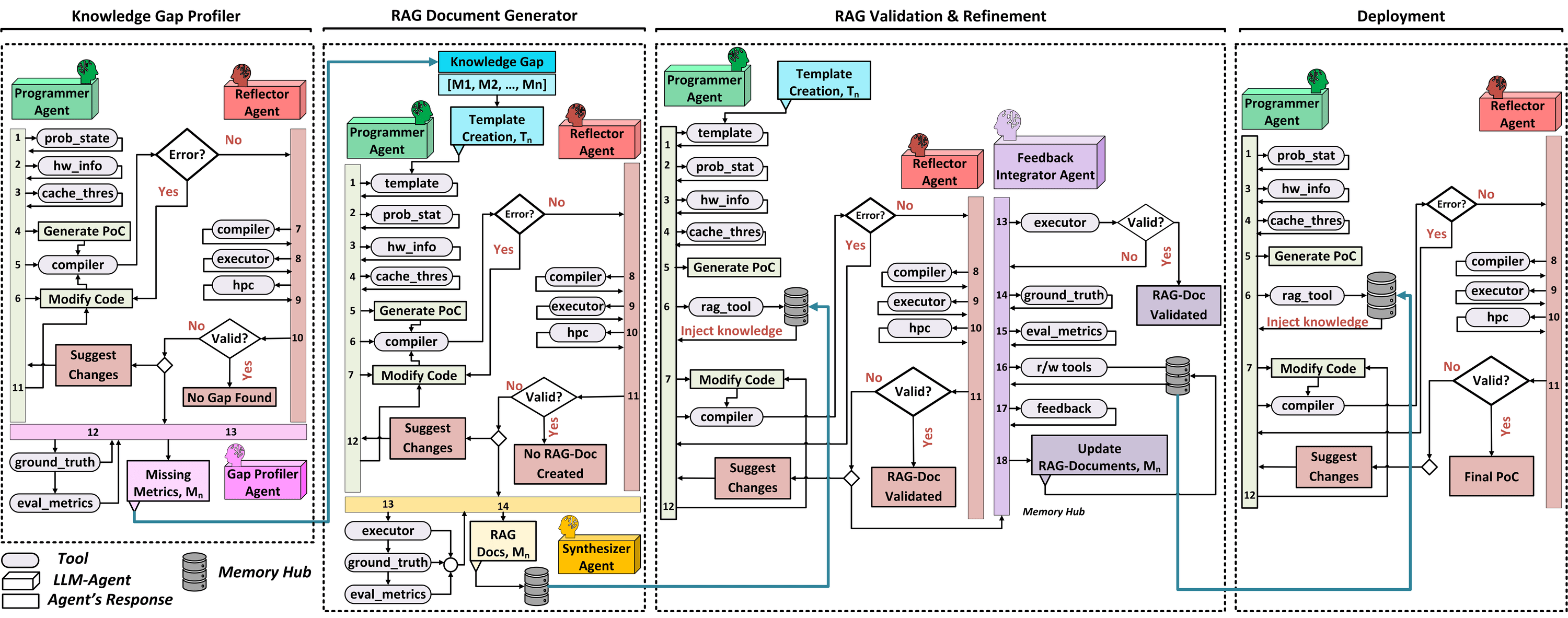}
    \caption{Multi-stage workflow of \ugen{}: S1 (Knowledge Gap Profiler) identifies overlooked, misgenerated, or misplaced attack attributes in \ac{llm}-generated \ac{poc} code; S2 (\ac{rag} Document Generator) generates domain-specific details for attack attributes; S3 (\ac{rag} Validation \& Refinement) validates and refines generated \ac{rag} documents; S4 (Deployment) refers to the final deployed  stage for the end-user. 
    }
    \label{fig:stage_all}
\end{figure*}

The first stage (S1) of \ugen systematically identifies where and why \acp{llm} fail to synthesize functionally correct microarchitectural attack \acp{poc} as illustrated in \autoref{fig:stage_all}. To isolate intrinsic reasoning limitations of \acp{llm}, we deliberately disable retrieval-augmented knowledge and require models to rely solely on their internal reasoning capabilities.

\paragraph{Problem statement construction} We begin by selecting a set of target attacks and supported victim functions. For each attack–victim pair, we manually craft a precise problem statement that specifies the threat model, victim semantics, and expected leakage behavior. This explicit specification constrains the problem space and prevents \acp{llm} from defaulting to generic attack templates. Consequently, the evaluation directly tests whether an \ac{llm} can synthesize victim-specific \acp{poc}, including correct adaptation of speculation triggers and cache behavior. \autoref{sec:problemstatement} shows an example problem statement for Spectre-v1.

\paragraph{Attack metric decomposition}
To localize failure modes, we decompose each attack into a set of attack metrics (c.f., \autoref{app:metrics}, \autoref{tab:offline-s1}) and analyze which metrics an \ac{llm} omits, misconstructs, or misplaces. An attack metric is a minimal, semantically meaningful building block that induces a required microarchitectural effect (e.\,g., controlled branch misprediction, cache set eviction, or high-resolution timing). This approach avoids penalizing benign syntactic variation while exposing semantically incorrect implementations that may appear superficially valid. For objective comparison, we annotate ground-truth \acp{poc} with explicit attack metrics using inline comments (\autoref{lst:code_ground_truth}). These annotations allow the Gap Profiler agent to compare generated and reference \acp{poc} based on microarchitectural functionality rather than syntax, enabling \ugen to surface genuine knowledge gaps and guide targeted augmentation in later stages.

\paragraph{Iterative \ac{poc} synthesis and validation} Given a target attack vector, programming language, and system-specific parameters, the agents initiate collaborative synthesis. The Programmer parses the problem statement, gathers relevant microarchitectural properties (e.\,g., cache size and associativity) and synthesizes a candidate \ac{poc} using its native \ac{llm} knowledge. The code is compiled and iteratively repaired until compilation succeeds. The Programmer then forwards the \ac{poc} to the Reflector, which evaluates exploitability using attack-specific validation tests and benchmarks. If validation fails, the Reflector returns concrete system-level feedback that the Programmer incorporates into subsequent refinement rounds. This process continues until the \ac{poc} satisfies validation criteria or reaches a predefined iteration limit.


\begin{standalonecode}[linebackgroundcolor=\lstHighlightLines{5}{9}]{Ground Truth for Spectre-v1 Metric M11}{lst:code_ground_truth}
int main() {
  size_t mal_x = (size_t)(secret - (char *)array1); 
  int i, score[2], Length = strlen(secret);
  uint8_t value[2];
  /* M11: Array/Probe Initialization */
  for (i = 0; i < sizeof(array2); i++){
    array2[i] = 1; 
  }
  /***********************************/
  printf("Reading %d bytes:\n", Length);
  while (--Length >= 0) {
    [...]
  }
  return (0);
}
\end{standalonecode}

\paragraph{Knowledge gap identification}
When validation repeatedly fails after 8 attempts, control transfers to the Gap Profiler, which compares the generated \ac{poc} against the ground-truth specification and evaluation criteria. Missing, incomplete, or misplaced attack metrics or metrics are labeled as \emph{Missing} and recorded in a per-run report within the \ac{poc} directory; otherwise, the Gap Profiler emits no output. We aggregate these reports across runs and compute the success rate for each metric (\autoref{fig:offline-s1}). These success rates provide a quantitative characterization of \ac{llm} deficiencies and explicitly identify which microarchitectural attack components require targeted knowledge augmentation via \ac{rag} in later stages, namely those with lower success rates.

\subsection{RAG Document Generator}

The second stage (S2) of \ugen addresses the knowledge gaps by reliably generating targeted retrieval-augmented guidance for attack metrics. Throughout this paper, we refer to this retrieval-augmented guidance as a \ac{rag} document. Rather than enriching the model with broad attack descriptions, we isolate individual attack metrics and systematically evaluate whether additional, structured guidance is required for correct synthesis. \autoref{fig:stage_all} illustrates the overall workflow of this stage, which is decomposed into the following steps:

\paragraph{Template Construction}
We first identify attack metrics that exhibit persistent reconstruction failures in S1. For each such metric, we derive a corresponding template by modifying the baseline \ac{poc} to omit exactly one attack attribute associated with that metric. 
This controlled omission isolates the reconstruction of a single metric and prevents confounding failures caused by unrelated components. Consequently, the evaluation directly assesses whether the \ac{llm} can recognize and reconstruct the omitted microarchitectural behavior using its internal knowledge and the provided ground-truth context.

\paragraph{Template-driven PoC reconstruction}
Given a target metric and its corresponding template, the Programmer Agent attempts to complete the \ac{poc} by reasoning over the problem specification, victim semantics, and hardware configuration. The agent operates under the same constraints as in S1, and relies exclusively on its native knowledge without retrieval augmentation. 
After each modification, the generated \ac{poc} is validated against attack-specific success criteria, including functional correctness and observable leakage behavior by the Reflector Agent. This process continues iteratively until the template is successfully patched or a predefined recursion limit is reached. The Programmer--Reflector interaction constitutes the testing phase shown in Figure~\ref{fig:architecture}(b), which operates identically across S1 through S3. The Synthesizer Agent is introduced at the end of S2 only when the testing phase fails to reconstruct a missing metric. 

\paragraph{Metric-specific RAG document synthesis} If the patched template passes validation, the metric is marked as self-recoverable and no \ac{rag} document is generated. 
If validation fails consistently, control transfers to the Synthesizer Agent, signaling that the missing metric represents a genuine knowledge gap that requires explicit augmentation.
In this scenario, the Synthesizer Agent analyzes the ground-truth implementation, the evaluation criteria, and the failed \ac{poc} attempts to identify the precise microarchitectural behavior that is missing or incorrectly implemented. The agent then generates a metric-specific \ac{rag} document following a predefined structure.

\paragraph{Design and structure of RAG documents} The predefined format of each \ac{rag} document remains consistent. Each \ac{rag} document provides: (1) a concise explanation of the attribute's significance in the context of microarchitectural attacks, (2) detailed step-by-step guidance for implementation, and (3) general placement instructions to ensure correct integration within the \ac{poc} code (\autoref{app:rag}, \autoref{lst:template}). This structure is intentionally designed to balance motivation with actionable guidance. We find that explicitly stating the importance of a metric is critical for successful synthesis, as \acp{llm} often disregard retrieved guidance that appears optional or non-essential. For example, in Spectre-v1, the controlled delay metric is frequently ignored unless its role in extending the speculative window is clearly emphasized; without this context, the model tends to omit the delay even when detailed implementation steps are provided.

\paragraph{Persistent knowledge storage}
At the end of this stage, the Synthesizer Agent stores all generated \ac{rag} documents in the memory hub, enabling persistent, metric-specific knowledge augmentation for subsequent runs. As a representative example, \autoref{app:rag}, \autoref{lst:rag-example} presents the \ac{rag} document for metric M11 (Array/Probe Initialization), which we identify as the most frequently missed attribute by GPT-4o (cf. \autoref{fig:offline-s1}). This example follows the predefined template and demonstrates how targeted, structured retrieval mitigates recurring \ac{llm} failure modes. 

\subsection{RAG Validation \& Refinement} \label{sec:methodology_s3}
The third stage (S3) empirically validates the effectiveness of the generated \ac{rag} documents and refines them through an iterative, feedback-driven process (\autoref{fig:stage_all}. At this stage, retrieval augmentation is enabled, and the Programmer Agent has access to the memory hub containing all metric-specific \ac{rag} documents generated during S2.

\paragraph{Targeted retrieval and metric-level patching}
For each template with an omitted attack metric, the Programmer Agent retrieves guidance for only the target metric and attempts to patch the template accordingly. The retrieval query is explicitly scoped to the missing metric, ensuring that the agent reasons about one metric at a time. This targeted retrieval strategy prevents context dilution and avoids interference from unrelated attack components. After applying the retrieved guidance, the patched \ac{poc} is compiled and evaluated against attack-specific success criteria.

\paragraph{Metric-level unit testing}
Each metric is validated independently through a unit-level test, where success is defined as the Programmer-Reflector loop (Testing Phase) successfully reconstructing the missing metric. A failure at this stage indicates that the retrieved \ac{rag} document is insufficient to guide correct synthesis of the target metric. This unit-testing abstraction allows \ugen localizing failures to individual attack components.

\paragraph{Feedback-driven RAG refinement}
When a unit test fails, control transfers to the Feedback Agent (FA), which initiates a refinement cycle for the corresponding \ac{rag} document. The FA analyzes the failed \ac{poc}, the retrieved guidance, and validation feedback to determine the root cause. 
The \ac{rag} document is then revised to improve clarity and precision, with emphasis on actionable implementation and placement guidance.

\paragraph{Expert-in-the-loop intervention}
If automated refinement fails to resolve the issue, a domain expert is introduced into the loop. The expert reviews the failed \ac{poc} and the current \ac{rag} document. Next, the expert provides targeted feedback, such as refining steps, clarifying constraints, or removing misleading details. This feedback is incorporated into the \ac{rag} document and stored back into the memory hub for subsequent validation. A representative example is shown in ~\autoref{appendix:expert_feedback} for metric M3 (Controlled Branch Misprediction), where expert feedback refines the interleaving logic to enforce branchless training and correct placement relative to the victim function.

\paragraph{Iterative convergence and finalization}
Each time a \ac{rag} document is modified, S3 requires to be re-executed to go through the unit testing. This refinement loop continues until the Programmer Agent successfully patches the template in a majority of trials, indicating that the \ac{rag} document reliably guides synthesis of the target metric. Once validated, the refined \ac{rag} document is finalized and made available for reuse in the deployment stage. 

\subsection{Deployment}\label{subsec:methodology_deployment}

The deployment stage (S4) represents the end-user-facing workflow of \ugen, where a non-expert user can generate a fully functional microarchitectural attack \ac{poc} without direct involvement in attack engineering or \ac{rag} document construction. This stage builds on the validated \ac{rag} documents and agent workflows established in earlier stages and exposes them through a controlled, metric-driven interface.

\paragraph{User-driven configuration and problem specification}
Deployment begins with the user specifying the attack vector, target programming language, and a problem statement describing the victim function and execution context. The problem statement allows the user to customize the victim logic, enabling \ugen to adapt the generated \ac{poc} to application-specific code without requiring the user to reason about low-level microarchitectural details. This design explicitly supports defender-oriented use cases, where the goal is to assess whether a given victim implementation is vulnerable under realistic conditions.

\paragraph{Initial PoC synthesis from scratch}
Given the user inputs, the Programmer Agent starts synthesizing a complete \ac{poc} from scratch without relying on any template. At this stage, the agent uses only its native reasoning and the problem specification to produce an initial draft of the attack code. This draft serves as the baseline into which retrieval-augmented knowledge will be incrementally integrated.

\paragraph{Sequential retrieval and augmentation}
The Programmer Agent processes the metric queries sequentially. For each query, it retrieves the corresponding metric-specific \ac{rag} document and examines whether the current \ac{poc} already reflects the retrieved guidance. If the \ac{poc} does not implement the required behavior correctly, the Programmer Agent patches the code using the retrieved implementation and placement guidance. Only after the agent confirms that the current metric is properly integrated it proceeds to the next query. If a user provides eight metric queries for Spectre-v1, the agent performs eight successive retrieval–verification–patch cycles, each targeting a single evaluation metric. This sequential augmentation strategy prevents context overload and ensures that each attack metric is incorporated correctly.

\paragraph{End-to-end validation and refinement}
Once all query-specific knowledge has been integrated and verified by the Programmer Agent, the final \ac{poc} is forwarded to the Reflector Agent for validation. The Reflector executes the \ac{poc} in the target environment and evaluates attack success using attack-specific validation criteria. If validation fails, the Programmer and Reflector Agents enter a bounded refinement loop, iteratively correcting errors until the \ac{poc} satisfies the success criteria or a predefined recursion limit is reached.

\paragraph{Practical deployment for non-expert users}
By allowing users to control attack construction through problem statements and metric-level queries, the deployment stage enables systematic vulnerability testing without requiring microarchitectural expertise. The combination of sequential retrieval, explicit verification of each metric, and automated validation ensures that generated \acp{poc} are functionally correct.

\section{Experiment Setup}
\label{sec:experimentalsetup}

\ugen is implemented using LangChain
(v1.2.6) and LangGraph
(v1.0.6) for orchestration. LangGraph is used to define the agent execution graph and enforce explicit control flow between code generation, reflection, retrieval, and evaluation stages. Each agent operates as an independent LangChain runnable with its own prompt template, bound tools, and access to a shared state. The shared state tracks conversation history, tool outputs, iteration counters, and retrieval status, enabling consistent refinement across iterations. 
All experiments are executed inside a Docker container based on Ubuntu 22.04. The evaluation is conducted on three hardware platforms: AMD Ryzen 9, Intel Tiger Lake i7-1165G7 @ 2.80GHz (Intel Tiger Lake), and ARM Neoverse-N1, all running Ubuntu 24.04 as the host OS.

\paragraph{\ac{llm}s}
We evaluate \ugen using three \acp{llm}: GPT-4o, Claude Sonnet-4, and Qwen3-Coder. GPT-4o is accessed via OpenAI's API 
based on the \texttt{gpt-4o} model, while Claude Sonnet-4 is accessed through Anthropic's native API
based on the \texttt{claude-sonnet-4-20250514} model. Qwen3-Coder is accessed through the Together AI platform
using the \texttt{Qwen3-Coder-480B-A35B-Instruct-FP8} model. 
Prompt files are organized according to the different stages of \ugen, reflecting the distinct objectives and constraints at each phase of the workflow.

\section{Evaluation}
\label{sec:eval}

Our evaluation is structured around five research questions:

\paragraph{RQ1}
To what extent can state-of-the-art \acp{llm} generate functionally correct and correctly placed microarchitectural attack metrics without retrieval augmentation? (\autoref{subsec:knowledge})

\paragraph{RQ2}
How effective are automatically generated \ac{rag} documents in enabling \ugen to correctly synthesize and insert missing attack metrics into \ac{poc} templates? (\autoref{subsec:rag})

\paragraph{RQ3}
Can retrieval-augmented \ugen autonomously generate fully functional microarchitectural attack \acp{poc} from scratch, and what are the resulting success, runtime, and cost tradeoffs across models? (\autoref{subsec:deployment})

\paragraph{RQ4} To what degree is \ugen portable across microarchitectures with varying configurations? (\autoref{subsec:portability})

\paragraph{RQ5} Can \ugen adapt to different vulnerable victim function variants and successfully generate exploitable Spectre-v1 \acp{poc} without manual re-engineering? (\autoref{subsec:victim})

The results in this section are obtained by running \ugen{} on AMD Ryzen 9.

\subsection{Knowledge Gap Analysis of LLM-Generated PoCs}\label{subsec:knowledge}
To answer RQ1, we quantify the knowledge gap of \acp{llm} using the \ugen framework. We first evaluate \ugen without any \ac{rag}-based knowledge augmentation during code generation. We focus on two representative and well-studied microarchitectural attacks, namely Prime+Probe and Spectre Variant 1, which expose complementary challenges. Prime+Probe requires correct eviction-set construction and precise timing analysis, whereas Spectre relies on speculative execution behavior and control-flow mistraining. We evaluate Claude Sonnet-4, GPT-4o, and Qwen3-Coder across all metrics by executing the Knowledge Gap Profiler ten times per model. \autoref{fig:offline-s1} summarizes the results.

\paragraph{Spectre} 
Our experiments reveal substantial variation across models in their ability to generate functional Spectre components. Claude Sonnet-4 performs the worst overall, failing to generate several critical metrics. In particular, the model is unable to construct the branch mistraining loop sequence (M4), resulting in a 0\% success rate. In contrast, Qwen3-Coder successfully generates all metrics in more than half of the runs. Its weakest point is controlled branch misprediction (M3), where the mistraining sequence is often incomplete. As a single non-functional metric prevents secret leakage, Qwen3-Coder fails to produce a fully working Spectre \ac{poc} in any run. Among the evaluated models, GPT-4o performs best overall: metrics M1, M2, and M7 are generated with 100\% success. The array/probe initialization (M11) metric is the most challenging one for GPT-4o, which limits the overall Spectre \ac{poc} code generation, resulting in a 2/10 success rate.

\paragraph{Prime+Probe}
Prime+Probe is more challenging for all three models, as it requires both eviction-set construction and a pointer-chasing linked-list setup. All models fail to generate a correct eviction set backed by a linked list, rendering the resulting attack snippets non-functional. Despite this limitation, all models achieve high success rates for cache contention detection and scenario sequencing metrics. However, GPT-4o lacks the ability to create eviction set creation for both attacker and victim, where Claude Sonnet-4 excels at generating successful eviction set creation.
In this attack, Claude Sonnet-4 achieves the highest accuracy for all the metrics, while GPT-4o fails to generate multiple metrics.

\begin{takeaway}
\textbf{Takeaway:} Current LLMs lack sufficient microarchitectural understanding to reliably generate end-to-end functional attack PoCs without external guidance, and their failure modes are model-specific rather than uniform. This heterogeneity motivates targeted RAG document generation tailored to the specific deficiencies of each LLM.
\end{takeaway}

\subsection{Effectiveness of Retrieval-Augmented Metric Synthesis}\label{subsec:rag}

For the Spectre \ac{poc} code, Claude requires five \ac{rag} documents without expert feedback to patch all the templates~\autoref{tab:template_metrics}.
Qwen3 requires six documents to be generated where one \ac{rag} document needed a revision from an expert.
The worst performance is obtained with GPT-4o, where eight documents are created of which five documents need expert feedback.
For the Prime+Probe attack, Claude only requires two \ac{rag} documents to be created to complete all templates.
Qwen3-Coder requires three documents, while GPT-4o needs four documents to create all metrics.
All models can generate \ac{rag} documents without expert feedback.
We observe that there are several metrics for both attacks, which require a \ac{rag} document for all models.

\newcommand{\ragdocsuccess}{{\footnotesize\faFile[regular]}}
\newcommand{\ragdocfailed}{{\footnotesize\xcancel{\faFile[regular]}}}
\newcommand{\ragdocsuccesswfeedback}{{\footnotesize\faFile*[regular]}}
\newcommand{\nofeedback}{{\footnotesize\faCircle[regular]}}
\newcommand{\feedback}{{\footnotesize\faCircle}}
\begin{table}[t]
    \footnotesize
    \centering
    \renewcommand{\arraystretch}{1.2} 
    \caption{
        Comparison of the completion capability of the tested \acp{llm} when tasked with a specific template for a certain metric (For AMD Ryzen 9).
        The programmer and reflector agents are either capable of filling the template successfully without (\ragdocfailed), after automatically generating a document (\ragdocsuccess), or after having the automatically generated document revised by an expert (\ragdocsuccesswfeedback). 
    }
    \label{tab:template_metrics}

\begin{tabular}{l rcl ccc}
\toprule
  & \multirow{2}{*}{\textbf{Metric}} & \multirow{2}{*}{$\rightarrow$} & \multirow{2}{*}{\textbf{Template}} & \textbf{Claude-} & \multirow{2}{*} {\textbf{GPT-4o}} & \textbf{Qwen3-} \\
  & & & & \textbf{Sonnet-4} & & \textbf{Coder} \\
\midrule
\multirow{11}{*}{\rot{Spectre-v1}}
    & M1       & $\rightarrow$ & T1  & {\ragdocfailed}  & {\ragdocfailed}           & {\ragdocfailed}           \\
    & M2       & $\rightarrow$ & T2  & {\ragdocfailed}  & {\ragdocfailed}           & {\ragdocfailed}           \\
    & M3       & $\rightarrow$ & T3  & {\ragdocsuccess} & {\ragdocsuccesswfeedback} & {\ragdocsuccesswfeedback} \\
    & M4       & $\rightarrow$ & T4  & {\ragdocfailed}  & {\ragdocsuccesswfeedback} & {\ragdocfailed}           \\
    & M5       & $\rightarrow$ & T5  & {\ragdocfailed}  & {\ragdocsuccesswfeedback} & {\ragdocfailed}           \\
    & M6       & $\rightarrow$ & T6  & {\ragdocsuccess} & {\ragdocsuccesswfeedback} & {\ragdocsuccess}          \\
    & M7       & $\rightarrow$ & T7  & {\ragdocsuccess} & {\ragdocsuccesswfeedback} & {\ragdocsuccess}          \\
    & M8\&M9   & $\rightarrow$ & T8  & {\ragdocsuccess} & {\ragdocsuccess}          & {\ragdocsuccess}          \\
    & M10      & $\rightarrow$ & T9  & {\ragdocfailed}  & {\ragdocsuccess}          & {\ragdocsuccess}          \\
    & M11      & $\rightarrow$ & T10 & {\ragdocsuccess} & {\ragdocsuccess}          & {\ragdocsuccess}          \\
    & M12      & $\rightarrow$ & T11 & {\ragdocfailed}  & {\ragdocfailed}           & {\ragdocfailed}           \\
\midrule
\multirow{7}{*}{\rot{Prime+Probe}}
    & M13\&M14 & $\rightarrow$ & T12 & {\ragdocfailed}  & {\ragdocsuccess}          & {\ragdocsuccess}          \\
    & M15      & $\rightarrow$ & T13 & {\ragdocsuccess} & {\ragdocsuccess}          & {\ragdocsuccess}          \\
    & M16      & $\rightarrow$ & T14 & {\ragdocfailed}  & {\ragdocfailed}           & {\ragdocfailed}           \\
    & M17      & $\rightarrow$ & T15 & {\ragdocfailed}  & {\ragdocsuccess}          & {\ragdocfailed}           \\
    & M18      & $\rightarrow$ & T16 & {\ragdocsuccess} & {\ragdocsuccess}          & {\ragdocsuccess}          \\
    & M19      & $\rightarrow$ & T17 & {\ragdocfailed}  & {\ragdocfailed}           & {\ragdocfailed}           \\
    & M20      & $\rightarrow$ & T18 & {\ragdocfailed}  & {\ragdocfailed}           & {\ragdocfailed}           \\
\bottomrule
                                                    
\end{tabular}
\end{table}

In practice, we observe that \ac{llm}-generated code often resemble baseline \acp{poc}, but with selective variation in specific attack components. For Spectre-v1, several metrics are frequently refactored by the models, particularly M3 (controlled branch misprediction), M5 (cache eviction targets \& placement), M8 (high-resolution timing \& serialization), and M10 (score accumulation \& early-stopping).
For example, the interleaving logic between branch mistraining and speculative access is sometimes reorganized, with variations in how training and attack iterations are mixed within loops.
Among the evaluated models, Claude Sonnet-4 and Qwen3-Coder exhibit more such refactorings than GPT-4o, which tends to preserve the baseline structure more closely.

In \autoref{tab:template_metrics}, T8 and T12 have been created by merging two metrics together.
T8 is formed by omitting both M8 (high-resolution timing \& serialization) and M9 (hit/miss classification threshold) for Spectre-v1 due to their close dependency.
As M8 and M9 are often implemented together as cache-based attack, merging them together improves the quality of the generated \ac{rag} document to provide a coherent implementation and placement guidance which we found to work better compared to creating separate documents for them.
A similar approach is taken for T12, where M13 and M14 are combined to create a single template.

The domain expert can choose three instructions--\texttt{ADD}, \texttt{REMOVE}, and \texttt{MODIFY}--to guide the FIA in upgrading the relevant \ac{rag} documents. The expert has access to the \ac{llm}-generated \ac{rag}-document and the \ac{poc} code.
By analyzing them, the expert determines which parts of the \ac{rag} document require updates, such as clarifying implementation steps or tightening placement constraints.
If the programmer agent fails to follow a specific step outlined in the \ac{rag} document, the expert can revise that step to improve clarity or add concrete examples to make the guidance easier for the \ac{llm} to apply. More detailed explanation with an example is given in \autoref{app:expert}. Every time when a \ac{rag} document is either created or refined, it is passed through unit testing for five times to observe the impact of the \ac{rag} document in patching the corresponding template \ac{poc}.
If the programmer agent can patch the relevant template code in four out of five runs, the \ac{rag} document is finalized for the deployment stage.

Our results demonstrate that automatically generated \ac{rag} documents are largely effective in closing metric-level knowledge gaps and enabling \ugen to synthesize missing attack components within \ac{poc} templates. In most cases, model-generated documents provide sufficient guidance for correct metric insertion without expert intervention, indicating that structured retrieval can substantially mitigate the limitations observed in the zero-augmentation setting. However, the degree of effectiveness varies across models: while Claude Sonnet-4 and Qwen3-Coder benefit from \ac{rag} with minimal refinement, GPT-4o frequently requires expert feedback to resolve ambiguities and tighten implementation constraints. This suggests that \ac{rag} effectiveness depends not only on document quality but also on how well an \ac{llm} absorbs external guidance, reinforcing the need for model-aware augmentation and validation strategies.

\begin{takeaway}
\textbf{Takeaway:} Automatically generated \ac{rag} documents are largely effective at enabling \ugen to correctly synthesize and integrate missing attack metrics into \ac{poc} templates. Claude Sonnet-4 succeeds without expert feedback across all evaluated metrics, demonstrating strong robustness to retrieval variability. In contrast, GPT-4o relies on expert feedback for multiple metrics, indicating greater sensitivity to underspecified or ambiguous retrieval guidance.
\end{takeaway}

\subsection{End-to-End PoC Generation Performance}\label{subsec:deployment}
In the deployment stage, the user utilizes \ugen to generate a functional \ac{poc} code.
The framework functions independently and fetches crucial information from available \ac{rag} documents.
During this phase, the programmer agent does not have any initial template to begin with; instead, it starts generating the code from scratch.

\paragraph{Success Rate} 
The success rate is computed over ten independent runs for each model and attack vector.
A run is considered successful if the generated \ac{poc} satisfies the attack-specific success condition defined in the problem statement.
For Spectre-v1, success requires leaking at least 50\% of the secret bytes correctly, where the secret is explicitly specified as part of the input problem statement.
For Prime+Probe, success is determined by observable cache contention (at least five CPU clock cycles) between the Prime$\rightarrow$Probe and Prime$\rightarrow$Victim$\rightarrow$Probe scenarios.

Each run is bounded by a recursion limit of 100 iterations, which caps the total number of node executions in the workflow, including agent invocations, tool executions, and retrieval steps.
The limit is identical for all attack vectors and models, providing a uniform upper bound on refinement effort while preventing non-terminating execution.

\autoref{table:eval_stage_4} summarizes the success rates of \ugen across different \acp{llm} and attack vectors during the deployment stage.
Overall, all models exhibit a clear improvement in success rate compared to knowledge gap profiling step, where \acp{llm} had no access to \ac{rag} documents. 
For instance, Claude Sonnet-4 achieves 90\% in generating fully functional \ac{poc} code for Spectre, confirming the effectiveness of retrieval-augmented, iterative refinement. GPT-4o still exhibits comparatively more failures for Spectre-v1.
We observe most of the failures attributed to minor errors in reconstructing M4 and M9.
Qwen3-Coder achieves a moderate success rate in generating Spectre \ac{poc} codes with a 70\% success rate.

For Prime+Probe, Qwen3-Coder achieves the highest success rate of 80\%, while both Claude Sonnet-4 and GPT-4o achieve 70\%.
In this attack, failures are more often caused by low-level implementation errors, such as segmentation fault while reconstructing M15. Additional failures stem from incorrect construction of the scenario sequencing (M19).
We observe that \acp{llm} sometimes encapsulate both prime and probe phases within a single helper function, which leads to an incorrect execution order and unsuccessful time difference.

\paragraph{Runtime Analysis} 
\autoref{table:eval_stage_4} reflects the average time required to generate a functional \ac{poc} code, averaged over ten runs per \ac{llm}.
Runtime in the deployment stage is primarily driven by recursive refinement behavior.
When validation fails, the agents engage in additional recursive steps for code revision.
Each recursion step involves an \ac{llm} agent invocation, tool execution, or a transition through the retriever node for \ac{rag}-based knowledge injection.
The number of recursive steps increases due to the following reasons: (1) code-level errors, such as compilation or execution errors, (2) reasoning errors related to missing, misordered, or incorrectly integrated attack metrics, and (3) the number of \ac{rag} documents that must be retrieved to complete a given attack.

\begin{table}[t]
    \footnotesize
    \centering
    \setlength{\tabcolsep}{2pt}
    \caption{Breakdown of success rate, cost, and runtime for the final deployment stage S4 (For AMD Ryzen 9).}
    \label{table:eval_stage_4}
    \begin{tabular}{l cc cc cc} 
        \toprule
        \multirow{2}{*}{\textbf{Model}}
                        & \multicolumn{2}{c}{\textbf{Success Rate (\%)}}
                                                             & \multicolumn{2}{c}{\textbf{Cost (\$)}}
                                                                                                  & \multicolumn{2}{c}{\textbf{Runtime (s)}} \\ 
                          \cmidrule(lr){2-3}                   \cmidrule(lr){4-5}                   \cmidrule(lr){6-7}
                        & \textbf{Spectre-v1} & \textbf{P+P} & \textbf{Spectre-v1} & \textbf{P+P} & \textbf{Spectre-v1} & \textbf{P+P}       \\ 
        \midrule
        Claude Sonnet-4 & 90                  & 70           & 3.01                & 1.57         & 351.1               & 223.1              \\
        GPT-4o          & 60                  & 70           & 2.30                & 1.15         & 280.8               & 252.0              \\
        Qwen3-Coder     & 70                  & 80           & 1.25                & 1.22         & 349.0               & 427.5              \\
        \bottomrule
    \end{tabular}
\end{table}

For Spectre-v1, both Qwen3-Coder and Claude Sonnet-4 require around 350 seconds on average to complete \ac{poc} generation.
Both models generate more verbose code and explanations compared to GPT-4o, including extensive comments, and debugging output.
Since agent responses are propagated across multiple nodes during recursive refinement, this verbosity contributes to increased runtime relative to GPT-4o.

The runtime for generating Prime+Probe takes less time for both Claude Sonnet-4 (223.1s) and GPT-4o (252.0s) compared to Spectre-v1.
Spectre-v1 generally incurs higher runtime than Prime+Probe due to its larger set of dependent metrics, which trigger more frequent transitions through the retriever node and cause additional validation cycles.
However, an exception is observed for Qwen3-Coder, which exhibits a higher runtime (427.5 s) for Prime+Probe due to the frequent compilation and execution errors, leading to additional recursive interactions between the programmer and reflector agents to fix the errors.

\paragraph{Cost Analysis} Cost is derived from the total number of tokens processed by each model, including both input tokens (system and agent prompts, retrieved \ac{rag} content, and tool outputs injected into the context) and output tokens (generated code, analysis, and feedback).
In addition, different \ac{llm} providers apply different pricing per million token usage, which directly affects the overall cost.

Among the evaluated models, Qwen3-Coder costs the least, primarily due to its lower per-million-token pricing compared to GPT-4o and Claude Sonnet-4.
In contrast, Claude Sonnet-4 incurs the highest cost since it generates more verbose outputs with additional comments on the generated code, resulting in higher output token consumption per invocation.

The \ac{rag} integration also contributes to the cost, as retrieved documents consume more input tokens for each agent.
Spectre-v1 \ac{poc} code generation task requires substantially more \ac{rag} documents, which is more expensive than Prime+Probe for GPT-4o and Claude Sonnet-4.
However, the cost difference between the two attack vectors is smaller for Qwen3-Coder since Qwen3-Coder undergoes more recursive refinement cycles for Prime+Probe, leading to additional agent invocations and higher cumulative token usage.

\begin{takeaway}
\textbf{Takeaway:} \ugen significantly improves the synthesis of fully functional \acp{poc} across all evaluated \acp{llm} by augmenting native model reasoning with targeted \ac{rag}. Attacks with higher microarchitectural complexity, such as Spectre-v1, incur increased runtime and cost due to deeper recursive refinement and more frequent retrieval.
\end{takeaway}
\begin{table}[t]
    \footnotesize
    \centering
    \caption{Evaluation of \acp{llm} for five Spectre-v1 victim function examples chosen from~\cite{kocher2018spectre} (For AMD Ryzen 9).}
    \label{tab:victim_function}
    \begin{tabular}{lccc}
        \toprule
        \multirow{2}{*}{\textbf{Victim Function}} & \multicolumn{3}{c}{\textbf{Success Rate (\%)}}                    \\
                                                    \cmidrule(lr){2-4} 
                                                  & \textbf{Claude Sonnet-4} & \textbf{GPT-4o} & \textbf{Qwen3-Coder} \\
        \midrule
        Example 4                                 & 60                       & 60              & 40                   \\ 
        Example 9                                 & 80                       & 60              & 60                   \\ 
        Example 12                                & 80                       & 80              & 80                   \\ 
        Example 14                                & 60                       & 60              & 60                   \\ 
        Example 15                                & 80                       & 60              & 80                   \\
        \bottomrule
    \end{tabular}
\end{table}

\subsection{Portability Across Microarchitectures}\label{subsec:portability}

Porting microarchitectural attack \acp{poc} across hardware platforms is a challenging task, as a working \ac{poc} depends on low-level ISA and microarchitecture specifications. The three platforms (Inter Tiger Lake, AMD Ryzen 9, and ARM Neoverse-N1) in our experimental setup differ in cache associativity, timing interfaces, cache maintenance instructions, serialization metrics, and cache hit/miss thresholds as summarized in Table~\ref{table:hw_diff}. Due to these differences, the portability of a \ac{poc} is unreliable without platform-specific calibration and code adaptation.

\begin{table}[t]
    \centering
    \footnotesize
    \setlength{\tabcolsep}{3pt}
    \caption{ISA and microarchitecture differences across Intel, AMD, and ARM processors required for successful generation of attacks.}
    \label{table:hw_diff}
    \begin{tabular}{lccc}
        \toprule
         & \textbf{Intel Tiger Lake} & \textbf{AMD Ryzen 9} & \textbf{ARM Neoverse-N1}\\
        \midrule
        Cache Assoc. & 12  & 8  & 4  \\
        Hit/Miss Thres. & 180-210   & 250-270  & 2 \\
        Timer Interface   & rdtscp  & rdtscp  & CNTVCT\_EL0  \\
        Flush Inst.  & clflush & clflush  & DC CIVAC   \\
        
        \bottomrule
    \end{tabular}
\end{table}

We evaluate the portability of \ugen{} on Intel Tiger Lake, AMD Ryzen 9, and ARM Neoverse-N1 with Spectre-v1 and Prime+Probe attacks. In our experiments, portability refers to \ugen{}'s ability to generate a functionally successful \ac{poc} on each target platform without manually rewriting the attack logic for that platform. \ugen{} supports the portability through three mechanisms. Microarchitecture-agnostic \ac{rag} documents describe attack requirements in a high-level to be leveraged in different ISAs and hardware configurations. The multi-agent framework iteratively detects and repairs missing or incorrect attack components. Finally, profiling tools measure platform-specific parameters such as timing thresholds, cache behavior, and eviction set requirements.

\ugen{} is executed ten times on each platform and \ac{llm} backend, and the results are summarized in Table \ref{table:portability}. For Spectre-v1, Claude Sonnet-4 achieves 100\% success rate on ARM and 90\% on AMD, 80\% on Intel, while GPT-4o and Qwen3-Coder achieve 60\% and 70\% success rate, respectively, on all three platforms. For Prime+Probe, all models perform similarly with 70-80\% success rate. Especially, the ARM results show that \ugen{} can utilize the \ac{rag} documents and microarchitecture profiling tools to construct a successful \ac{poc} code tailored for AArch64-specific timing, cache maintenance, and serialization mechanisms. Across both attacks, most failures are due to reasoning and implementation errors in the generated \ac{poc} code. This outcome indicates that \ugen{}'s profiling tools and \ac{rag} components provide accurate attack and platform-specific information, while the remaining portability limitations primarily arise from the \ac{llm} backend's ability to correctly integrate that information into the executable attack code. 

\begin{table}[t]
    \centering
    \footnotesize
    \setlength{\tabcolsep}{3pt}
    \caption{Portability of \ugen{} across CPU (micro-) architectures. Success rates (\%) are reported for two attack scenarios (Spectre-v1 and Prime+Probe) across three \ac{llm} backends.}
    \label{table:portability}
    \begin{tabular}{ l ccc ccc }
        \toprule
        \multirow{2}{*}{\textbf{Processor}} & \multicolumn{3}{c}{\textbf{Spectre-v1}} & \multicolumn{3}{c}{\textbf{Prime+Probe}} \\
        \cmidrule(lr){2-4} \cmidrule(lr){5-7}
         & \textbf{Claude} & \textbf{GPT} & \textbf{Qwen} & \textbf{Claude} & \textbf{GPT} & \textbf{Qwen} \\
        \midrule
        Intel Tiger Lake & 80  & 60  & 70  & 80  & 70  & 80  \\
        AMD Ryzen 9      & 90  & 60  & 70  & 70  & 70  & 80  \\
        ARM Neoverse-N1  & 100 & 60  & 70  & 70  & 70  & 70  \\
        \bottomrule
    \end{tabular}
\end{table}

\begin{takeaway}
\textbf{Takeaway:} \ugen generalizes across diverse microarchitectures, including cross-ISA portability to AArch64, by utilizing the provided tools with \ac{rag}-guided attack adaptation. While Claude provides the best performance, GPT struggles with successful \ac{poc} code generation more frequently.
\end{takeaway}

\subsection{Generalization Across Victim Function Variants}\label{subsec:victim}
Here, our goal is to observe \ugen{'s} adaptability to different variants of the vulnerable victim functions.
Paul Kocher~\cite{kocher2018spectre} released 15 victim functions vulnerable to Spectre-v1.
We choose the five victim functions that require modifications to the ground truth Spectre \ac{poc} to leak the secret.
If a victim function example can be exploited without making any changes on the \ac{poc}, we exclude it for the experiment. 

The victim function is included in the problem statement that serves as input to the programmer agent.
The performance of different \acp{llm} with five victim function variants are listed in \autoref{tab:victim_function} computed over five runs.
Among the evaluated \acp{llm}, Claude Sonnet-4 achieves the highest functional \ac{poc} code generation rate in the range of 60\%-80\%.
GPT-4o and Qwen3-Coder have similar performance on generating successful \ac{poc} codes.
This experiment is crucial for different use cases where a user analyzes the vulnerabilities of a custom function.
\ugen can generate functional \ac{poc} codes based on the given victim function, demonstrating the re-usability of the generated \ac{rag} documents with advanced reasoning. 

\begin{takeaway}
\textbf{Takeaway:} \ugen can successfully transfer the retrieved knowledge from \ac{rag} documents in generating \ac{poc} codes when the target victim function is altered by the user.
\end{takeaway}

\section{Discussion}
\label{sec:discussion}

\paragraph{Open Source \acp{llm} Usage}
Our evaluation shows that the effectiveness of open-source \acp{llm} in \ugen is primarily determined by their compatibility with agentic tool invocation rather than by model openness itself. Several open-source models we explored, such as DeepSeek and LLaMA-based variants, exhibited frequent failures in tool interaction when integrated through LangChain, including dropped tool calls, malformed arguments, or incomplete execution traces. These limitations prevent reliable compilation, execution, and validation of generated code, making them unsuitable for end-to-end microarchitectural \ac{poc} synthesis in our framework. In contrast, Qwen3-Coder demonstrates that open-source models can perform competitively when exposed through a stable OpenAI-compatible API with robust tool-calling support. 

\paragraph{Jailbreaking}
Some \acp{llm} do not allow users to produce a working attack code. Instead, they provide a pseudocode consisting of several metrics to construct an attack code. While it is known that jailbreaking techniques can be employed to bypass such barriers~\cite{DBLP:journals/corr/abs-2410-15236}, it is out of scope for our paper. 

\paragraph{Usage of Cross-RAG Documents}
\ac{rag} documents generated by an \ac{llm} can also be utilized for another \ac{llm}. We evaluated the usage of GPT-4o created \ac{rag} documents with the Claude and Qwen programmer agents. However, each \ac{llm} has a different way of reasoning based on the instructions in the \ac{rag} documents. Hence, the success rate significantly drops when \ac{llm} agents are fed with \ac{rag} documents generated by another \ac{llm}.

\section{Related Work}
In this section, we provide an overview of current applications of \acp{llm} for offensive \ac{poc} code generation, as well as multi-agent frameworks and \ac{rag} integration in cybersecurity.

\paragraph{PoC Code Generation with LLMs}
Automating \ac{poc} exploit generation is a challenging task due to incomplete vulnerability context and the absence of executable environment details. Prior work has explored the use of \acp{llm} to mitigate these limitations. PoCGen~\cite{DBLP:journals/corr/abs-2506-04962} autonomously generates and validates \ac{poc} exploits for vulnerabilities in npm packages using multiple \ac{llm} agents responsible for vulnerability analysis, exploit generation, and validation. Fang et al.~\cite{DBLP:journals/corr/abs-2404-08144} showed that one-day vulnerabilities can be exploited by a single GPT-4 model with a high success rate using Common Vulnerabilities and Exposures (CVE) reports. Similarly, Zhao et al.~\cite{DBLP:journals/corr/abs-2510-10148} further evaluate the reasoning capabilities of GPT-4o and DeepSeek-R1 for \ac{poc} generation across a curated set of CVEs. Ullah et al.~\cite{DBLP:journals/corr/abs-2509-01835} extend single-\ac{llm} approaches by proposing a multi-agent framework that extracts key information from CVEs to reproduce exploitation code.

\paragraph{Multi-Agent LLM Frameworks for Offensive Usage}
Single-agent \ac{llm} frameworks often struggle with multistep tasks due to limited iterative feedback and coordination. To address these limitations, prior work has adopted multi-agent architectures in which each agent performs a specialized subtask. \ac{llm}-based penetration testing systems commonly employ planner, code generator, verifier, debugger, and evaluator agents~\cite{mei2025autopen,DBLP:conf/asiaccs/Shen0LCZSWR25,DBLP:conf/trustcom/GinigeNJS25}. MalGEN~\cite{DBLP:journals/corr/abs-2506-07586} decomposes high-level malicious codes into sequences of smaller tasks, which are then assembled into fully functional malware artifacts. Several studies also examined the usage of multi-agent systems for Capture-the-Flag (CTF) challenges. D-CIPHER~\cite{DBLP:journals/corr/abs-2502-10931} utilizes prompter, planner, and executor agents to adapt the multi-agent system to different CTF challenges. CRAKEN~\cite{DBLP:journals/corr/abs-2505-17107} augments a planner–executor architecture with \ac{rag} to improve CTF-solving performance. Similarly, HackSynth~\cite{DBLP:journals/corr/abs-2412-01778} introduces a planner–summarizer multi-agent framework and proposes two new CTF benchmarks. 

\paragraph{RAG Integration into LLMs}
\acp{llm} often suffer from incomplete or outdated knowledge, which limits their effectiveness in offensive cybersecurity tasks. \ac{rag} augments \acp{llm} with external documents, enabling access to vulnerability details, domain concepts, and code artifacts. CRAKEN~\cite{DBLP:journals/corr/abs-2505-17107} integrates Self-\ac{rag} and Graph-\ac{rag} to support iterative retrieval, generation, refinement, and structured reasoning for cybersecurity challenges. PentestAgent~\cite{DBLP:conf/asiaccs/Shen0LCZSWR25} addresses inter-agent short-term memory limitations by incorporating environmental context and providing agents with curated reconnaissance tools for effective information gathering. AutoPentester~\cite{DBLP:conf/trustcom/GinigeNJS25} further equips generator agents with \ac{rag}-based tool documentation to produce complete, executable commands without human intervention.

\section{Conclusion}
\label{sec:conclusion}

We present \ugen, a \ac{rag}-empowered multi-agent framework and show that \ugen significantly enhances \acp{llm}'s capability of generating \ac{poc} code for microarchitectural attacks.
The evaluated \acp{llm} natively achieve success rates of only up to 20\%. \ugen increases the success rate up to 90\% for certain models.
Thereby, \ugen serves as a useful framework for the security community and system administrators as it enables researchers and practitioners to efficiently explore, adapt, and evaluate microarchitectural attacks across multiple microarchitectures, languages, and deployment settings.

\appendix
\section*{Ethical Considerations}
\paragraph{Innovations with both positive and negative potential outcomes}
As of today, microarchitectural attacks are considered to be rather sophisticated as they require expert knowledge, a deep understanding of the underlying hardware, and a rather long preparation phase.
However, we have seen machine learning and artificial intelligence improving boosting the development of all kinds of applications, including phishing campaigns and malware.
We estimate that malware developers will make use of \acp{llm} to lower the hurdle of successfully implementing microarchitectural attacks and thereby enrich their malware portfolio with such attacks in the future.
Therefore, it becomes increasingly important for defenders to thoroughly test the vulnerability of their systems for microarchitectural attacks.
We consider \ugen more helpful for defenders than attackers because it allows for generating benign \acp{poc} such that defenders can evaluate their systems \emph{today} while our success rates suggest that attackers may require a lot more work to adopt \ugen and use it for building actual exploits in the \emph{future}.
Thereby, while of course posing the dual-use question, we argue that \ugen helps defenders getting one step ahead of attackers while not meaningfully improving attacker's capabilities.

\paragraph{Terms of service}
While the usage policies of commercial \ac{llm} providers restrict the generation of malicious code, \ugen falls under controlled security research and defensive evaluation. The goal of this study is not to deploy, operationalize, or scale attacks, but to measure and understand the risks of \ac{llm}-assisted code generation for microarchitectural attack. All experiments are conducted in isolated and author-controlled environments on test hardware, and no production systems. Moreover, the generated \ac{poc} codes are well-known attacks in the literature. \ugen does not generate unknown new attacks that could be leveraged by malicious actors. We follow established norms in computer security research, where potentially harmful techniques are studied under controlled conditions to inform mitigations, guide system hardening, and support responsible risk assessment. We believe that empirically evaluating real-world \acp{llm} is necessary to obtain ecologically valid insights, as training or hosting comparable models locally is infeasible for most academic researchers. In this context, the benefits of understanding and mitigating these risks outweigh the limited and carefully contained exposure involved in our study.

\paragraph{Experiments with live systems}
All our evaluations took place in a controlled lab environment on-site.
The \ac{rag} system and its generated \acp{poc} were executed on local machines that were prepared solely for our tests as described in \autoref{subsec:portability}.
Only the \ac{llm} inference took place on cloud servers provided by the respective model providers.
Our interaction with the models strictly followed documented API description.
Generated \acp{poc} never targeted any publicly accessible live systems nor any systems that were shared with other parties.

\paragraph{Disclosures}
We did not discover new vulnerabilities and therefore did not start any disclosure process. All \acp{llm} are tasked clearly to create a \ac{poc} code without using any jailbreaking techniques.

\section*{Open Science}
The \ugen source code and \ac{rag} documents are available at \url{https://github.com/anonymoushei4444/uGEN.git}.

\bibliographystyle{plain}
\bibliography{Ref}


\section*{Tool Descriptions for Agents}\label{app:tools}

\ugen integrates a set of tools that connect \ac{llm} agents to the underlying system and execution environment. These tools are exposed through LangChain's tool interface and return structured outputs that directly influence agent decisions.

\subsection{Generic Tools}   
    \begin{description}[noitemsep,topsep=0pt,leftmargin=*]
      \item[\texttt{compiler}:] Compilation is performed using \texttt{gcc} for C-based PoCs and \texttt{g++} for C++ implementations. Rust-based PoCs are also supported via cargo, with the Rust toolchain installed inside the container using rustup. Generated source files are written to a run-specific working directory and compiled using standard command-line invocations.
    
       \item[\texttt{executor}:] Program execution is handled by a dedicated execution tool that enumerates compiled binaries and runs them under CPU pinning using \texttt{taskset}. Each binary is executed on a fixed CPU core with a bounded timeout error. Execution outputs and errors are captured and returned to the agents for validation.
    \end{description}
    
  \ugen includes other supporting tools for file-system operations, such as reading ground-truth PoCs and evaluation metrics, as well as storing and retrieving \ac{rag} documents and intermediate PoCs. These tools enable persistent state management across iterations and ensure that the tool outputs are consistently used by the agents throughout the workflow.

\subsection{Specialized System-level Tools}
  \begin{description}[noitemsep,topsep=0pt,leftmargin=*]
   \item[\texttt{hpc}:] Hardware Performance Counter (HPC) measurements are collected using Linux \texttt{perf}, which is built from the Linux stable kernel repository and installed directly inside the container. The \texttt{perf stat} interface is used to collect hardware performance counters relevant to side-channel analysis, with results parsed and returned to the framework. For each attack vector, the specific set of counters to be measured is selected by the \ac{llm} agent and passed to the tool as arguments, allowing attack-aware and targeted performance monitoring.

    \item[\texttt{hw\_info}:] System cache parameters are extracted using a cache information tool that invokes standard Linux interfaces, including \texttt{getconf -a}, \texttt{lscpu}, and sysfs entries under \texttt{/sys/devices/system/cpu/cpu0/cache}. These parameters are made available to agents during code generation to ensure hardware-aware PoC construction.

    \item[\texttt{cache\_thres}:] The cache hit/miss latency boundary used to distinguish cached from uncached accesses is calibrated at runtime rather than hard-coded, since it varies across microarchitectures and timer sources. The tool compiles a small C calibration program with \texttt{gcc}, selects an architecture-appropriate timer and flush instruction, measures the median latency of cached and flushed accesses under CPU pinning, and returns a recommended threshold biased toward the hit-side. The result is returned to the agent as a structured summary and embedded into the generated PoC to support hit/miss classification (metrics M9 and M18).

\end{description}

\section*{Evaluation Metrics for Attacks}\label{app:metrics}
\autoref{tab:offline-s1} lists all identified evaluation metrics for Spectre-v1 and Prime+Probe.

\begin{table}[t]
    \footnotesize
    \centering
    \caption{Evaluation Metrics for Spectre-v1 and Prime+Probe attacks.}
    \label{tab:offline-s1}
    \pgfplotstableread{Data/successrate-problemstatement.csv}{\data}
\pgfplotstabletypeset[
    columns={metric,desc},
    every head row/.style={
        before row={
            \toprule
        },
        after row={
            \midrule      
        },
    },
    every last row/.style={
        after row={
            \bottomrule
            \multicolumn{2}{l}{\footnotesize{M1-M12 refer to Spectre-v1, and M13-M20 refer to Prime+Probe.}}
        }
    },
    every nth row={12}{before row=\midrule},
    columns/metric/.style={
        string type,
        column type={l},
        column name={Metric},
    },
    columns/desc/.style={
        string type,
        column type={l},
        column name={Description},
    },
]{\data}

\end{table}

\section*{Knowledge Gap}
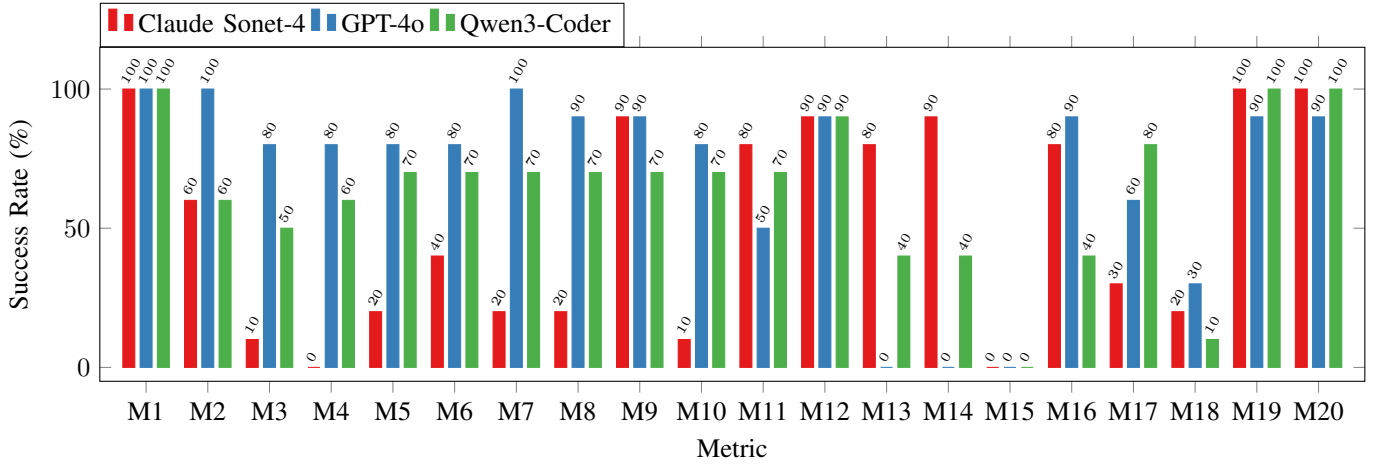
\begin{figure*}[t]
    \centering
    \pgfplotstableread{Data/successrate-problemstatement.csv}{\data}
\begin{tikzpicture}[
        plot/.style={draw,fill,text=black},
    ]
    \begin{axis}[
        height=6cm,
        width=1.01\linewidth,
        ybar=2pt,
        bar width=4.5pt,
        legend style={
            at={(0,1)},
            anchor=south west,
            legend columns=-1,
        },
        nodes near coords,
        nodes near coords align={south west},
        nodes near coords style={rotate=60,inner sep=1pt,font=\tiny},
        xmin=-.75,xmax=19.75,
        xtick=data,
        xticklabels from table={\data}{metric},
        xlabel={Metric},
        ymin=-5,ymax=115,
        ylabel={Success Rate (\%)},
    ]

        \addplot+[plot] table[x expr=\coordindex,y=sonet4w    ] {\data};
        \addlegendentry{Claude Sonet-4}


        \addplot+[plot] table[x expr=\coordindex,y=gpt4ow     ] {\data};
        \addlegendentry{GPT-4o}

        \addplot+[plot] table[x expr=\coordindex,y=qwen3coderw] {\data};
        \addlegendentry{Qwen3-Coder}
    \end{axis}
\end{tikzpicture}
    \caption{Knowledge Gap Profiler results showing the success rate of correctly implementing each metric in \ac{poc} codes across three \ac{llm} backends, evaluated over ten independent runs per model in the absence of retrieval augmentation.}
    \label{fig:offline-s1}
\end{figure*}

\autoref{fig:offline-s1} visualizes the success rate of S1 in generating the identified metrics over ten independent runs per model in the absence of retrieval augmentation.
Success rates of 80-100\% indicate a solid understanding.
Metrics that are only implemented correctly in 0--20\% of runs pose a severe road block when generating \ac{poc} code as chances are low that all of these metrics are implemented correctly in the same run, which is a requirement to achieve a successful \ac{poc}.

\section*{Expert Feedback with Example}\label{app:expert}

\autoref{appendix:expert_feedback} gives a representative example of how expert feedback refines the \ac{rag} document for the M3 (controlled branch misprediction) metric.
In this case, the initial \ac{rag} document often led the \ac{llm} to implement the interleaving logic using explicit branch conditions and to place it immediately before the victim function, which interferes with the attacker-controlled speculative execution and renders the attack ineffective.
In other instances, the \ac{llm} placed the interleaving logic inside the victim function, an approach that violates the threat model because the attacker should not assume control over victim code.
The expert feedback addresses these issues by explicitly stating that the interleaving logic must be branchless and by adding precise placement constraints, supported by examples.
These refinements significantly improve the \ac{llm}'s ability to reproduce the intended metric correctly.
However, adding more detail to a \ac{rag} document does not always improve \ac{llm} performance.
We observe that \ac{rag} documents generated by the synthesizer agent can sometimes become overly verbose, e.\,g., by including lengthy explanations of a primitive's importance or redundant elaboration of intermediate steps.
While such detail may be informative, it can also trigger a ``forgetting'' effect, where the \ac{llm} fails to follow all required implementation or placement steps consistently.
To address this, the expert can use \texttt{REMOVE} feedback to delete unnecessary content, such as redundant explanations or overly detailed background discussion in the ``Importance'' feedback section, and thereby keep the document focused on concise, actionable implementation and placement guidance.

\section*{Expert Feedback (S3)} \label{appendix:expert_feedback}
        \begin{tcolorbox}[title=\textbf{LLM-Generated RAG Document},
            colback=gray!3, colframe=gray!60, sharpish corners,
            fonttitle=\bfseries, coltitle=black, breakable, left=3pt, right=3pt]
            \footnotesize
            \textbf{Title: Controlled Branch Misprediction} \\
            
            \textbf{Importance:}
            
            Controlled branch misprediction [...] \\ 
            
            \textbf{Implementation Guidance:} To implement controlled branch misprediction, follow these steps:
            \begin{enumerate}
                \item Identify the Conditional Branch: Locate [...]
                \item Craft Misprediction Conditions: Design [...]
                \item Interleave with Legitimate Accesses: Mix legitimate accesses with speculative ones to train the branch predictor. This increases the likelihood of misprediction when the speculative path is executed.
                \item Ensure Speculative Execution: Use [...]
            \end{enumerate}
            
            \textbf{Placement Guidance:}
            Insert the controlled branch misprediction logic within the loop that prepares the speculative execution environment. This should [...]
        \end{tcolorbox}
        \begin{tcolorbox}[title=\textbf{Expert Feedback},
            colback=blue!3, colframe=gray!60, sharpish corners,
            fonttitle=\bfseries, coltitle=black, breakable, left=3pt, right=3pt]
            \footnotesize
            \textbf{ADD the following details under the Implementation Guidelines:}
            \begin{itemize}
              \item Interleave safe and malicious index values within the same loop.
              \item Use branchless arithmetic expressions; never use branching constructs such as \texttt{if} statements or ternary operators.
              \item Example: \texttt{index = a + cond * (b - a);} where \texttt{cond = !(j \% 6)} toggles between 0 and 1.
            \end{itemize}
            
            \textbf{ADD the following details under the Placement Guidelines:}
            \begin{itemize}
              \item Insert this interleaving logic before the index is used as the input to a victim function.
              \item This step must not be inside the victim function, as the attacker should not have access to modify it.
            \end{itemize}
        \end{tcolorbox}
        \begin{tcolorbox}[title=\textbf{RAG Document Updated by LLM},
            colback=green!3, colframe=gray!60, sharpish corners,
            fonttitle=\bfseries, coltitle=black, breakable, left=3pt, right=3pt]
            \footnotesize
            \textbf{Title: Controlled Branch Misprediction}\\
            
            \textbf{Importance:}
            
            Controlled branch misprediction [...]\\
            
            \textbf{Implementation Guidance:} To implement controlled branch misprediction, follow these steps:
            \begin{enumerate}
                \item Identify the Conditional Branch: Locate [...]
                \item Interleave with Legitimate Accesses: Mix legitimate accesses with speculative ones to train the branch predictor. This increases the likelihood of misprediction when the speculative path is executed.
                \begin{itemize}
                      \item Interleave safe and malicious index values within the same loop.
                      \item Use branchless arithmetic expressions; never use branching constructs such as \texttt{if} statements or ternary operators.
                      \item Example: \texttt{index = a + cond * (b - a);} where \texttt{cond = !(j \% 6)} toggles between 0 and 1.
                \end{itemize}
                \item Ensure Speculative Execution: Use [...]
            \end{enumerate}
            
            \textbf{Placement Guidance:}
            Insert the controlled branch misprediction logic within the loop that prepares the speculative execution environment. This should [...]
            \begin{itemize}
              \item Insert this interleaving logic before the index is used as the input to a victim function.
              \item This step must not be inside the victim function, as the attacker should not have access to modify it.
            \end{itemize}
        \end{tcolorbox}


\section*{RAG Document Template}\label{app:rag}

\begin{boxK}{Template for \ac{rag} Document}{lst:template}
\textbf{Title:} Name the targeted \texttt{attack\_attribute}

\textbf{Importance:} Explain why this component is essential for microarchitectural or side-channel attacks, and how it impacts the reliability or correctness of the \ac{poc}.

\textbf{Implementation Guidance:} Provide a detailed, step-by-step approach for generating the required patch or logic (consider the ground truth as reference). Use reasoning that is easy for any \acp{llm} to follow, and avoid code-specific details like variable names or syntax. You can give examples (such as equations, logics, etc.) if necessary.

\textbf{Placement Guidance:} Explain how to find the correct location in the code to insert the patch, using general principles (e.g., after measurement loops, before output assignment).
\end{boxK}


\begin{boxK}{An Example Document for M11}{lst:rag-example}
\textbf{{Title:}} Array/Probe Initialization

\textbf{Importance:}
Array or probe initialization is a critical step in microarchitectural attacks like Spectre-v1. It ensures that the cache lines are in a predictable state before the attack is executed. Without proper initialization, the cache behavior can be inconsistent, leading to unreliable results. This step is essential for creating a controlled environment where the attack can accurately measure cache hits and misses.

\textbf{Implementation Guidance:}
To implement array/probe initialization, follow these steps:
\begin{itemize}
    \item Initialize the array or probe with a specific value (e.g., all elements set to 1) before starting the attack loop.
    
    \item Ensure that the initialization covers the entire array or probe to avoid any uninitialized memory access.

    \item  This initialization should be done in the main function before entering the attack loop.

\end{itemize}

\textbf{Placement Guidance:}
The initialization should be placed at the beginning of the main function, right after declaring the variables and before entering the attack loop. This ensures that the array or probe is in the desired state before any speculative execution or cache timing measurements are performed.
\end{boxK}


\section*{Problem Statement}
\label{sec:problemstatement}
\begin{greenprompt}
\footnotesize
\textbf{Spectre Variant 1 (Spectre-v1) Proof-of-Concept Generation}

\medskip
\textbf{Problem Statement}

Generate a proof-of-concept program that demonstrates the Spectre Variant 1 (Spectre-v1) attack by leveraging speculative execution to leak a fixed secret string from memory using a microarchitectural side channel based on cache timing analysis.

\medskip
\textbf{Requirements}
\begin{itemize}[leftmargin=*, itemsep=1pt, topsep=1pt]
  \item \textbf{Secret Data}
    \begin{itemize}[leftmargin=1.2em, itemsep=1pt, topsep=1pt]
      \item The secret must be a fixed null-terminated string, for example:
\begin{promptlisting}
const char *secret = "It's a secret!!";
\end{promptlisting}
      \item The goal is to leak the content of this string byte by byte without accessing it directly through architectural control flow.
    \end{itemize}

  \item \textbf{Victim Buffer and Bounds Variable}
    \begin{itemize}[leftmargin=1.2em, itemsep=1pt, topsep=1pt]
      \item Define an array (\texttt{array1}) of size \texttt{L}:
\begin{promptlisting}
unsigned int array1_size = L;
uint8_t array1[L] = {0};
\end{promptlisting}
      \item The PoC should treat this array as a safe boundary for access.
    \end{itemize}

  \item \textbf{Victim Function (Fixed Form)}
    \begin{itemize}[leftmargin=1.2em, itemsep=1pt, topsep=1pt]
      \item The following function must be present as-is and used to trigger the speculative execution vulnerability:
\begin{promptlisting}
void victim_function(size_t x) {
    if (x < array1_size) {
        temp &= array2[array1[x] * 512];
    }
}
\end{promptlisting}
      \item Do not modify this function.
      \item The attacker is expected to generate surrounding logic to exploit this function via out-of-bounds speculative access.
    \end{itemize}

  \item \textbf{Leakage Target}
    \begin{itemize}[leftmargin=1.2em, itemsep=1pt, topsep=1pt]
      \item Your objective is to recover the bytes of the secret variable without reading it directly.
      \item Print the leaked bytes to standard output, including memory offset information and a confidence indicator for each guessed byte.
    \end{itemize}

  \item \textbf{Output Format}
    \begin{itemize}[leftmargin=1.2em, itemsep=1pt, topsep=1pt]
      \item For each leaked byte, print its memory offset, the guessed value (both as a hex value and as the corresponding ASCII character), and a confidence score.
    \end{itemize}

    \item \textbf{Expected Output}
    \begin{itemize}[leftmargin=1.2em, itemsep=1pt, topsep=1pt]
      \item If the PoC can leak at least half of the secret bytes correctly, consider it as a successful attack with expected behavior. 
    \end{itemize}
    
\end{itemize}
\end{greenprompt}

\begin{greenprompt}
\footnotesize
\textbf{Spectre Variant 1 (Spectre-v1) Proof-of-Concept Generation}

\medskip
\textbf{Problem Statement 2}

Generate a proof-of-concept program that demonstrates the Spectre Variant 1 [...] 

\medskip
\textbf{Requirements} 

\begin{itemize}[leftmargin=*, itemsep=1pt, topsep=1pt]
  \item ....
  \item \textbf{Victim Function V4(Fixed Form)}
    \begin{itemize}[leftmargin=1.2em, itemsep=1pt, topsep=1pt]
      \item The following function must be present as-is and used to trigger the speculative execution vulnerability:
\begin{promptlisting}
void victim_function_v4(size_t x) {
        if (x < array1_size) {
            temp &= array2[array1[x] * 512];
        }
    }
\end{promptlisting}
      \item Do not modify this function.
      \item The attacker is expected to generate surrounding logic to exploit this function via out-of-bounds speculative access.
    \end{itemize}

....
\end{itemize}

\end{greenprompt}

\begin{greenprompt}
\footnotesize
\textbf{Spectre Variant 1 (Spectre-v1) Proof-of-Concept Generation}

\medskip
\textbf{Problem Statement 3}

Generate a proof-of-concept program that demonstrates the Spectre Variant 1 [...] 

\medskip
\textbf{Requirements} 

\begin{itemize}[leftmargin=*, itemsep=1pt, topsep=1pt]
  \item ....
  \item \textbf{Victim Function V9(Fixed Form)}
    \begin{itemize}[leftmargin=1.2em, itemsep=1pt, topsep=1pt]
      \item The following function must be present as-is and used to trigger the speculative execution vulnerability:
\begin{promptlisting}
void victim_function_v9(size_t x, int *x_is_safe) {
                if (*x_is_safe)
                    temp &= array2[array1[x] * 512];
            }
\end{promptlisting}
      \item Do not modify this function.
      \item The attacker is expected to generate surrounding logic to exploit this function via out-of-bounds speculative access.
    \end{itemize}

....
\end{itemize}

\end{greenprompt}

\end{document}